\documentclass[10pt, twocolumn, comsoc]{IEEEtran}

\usepackage{graphicx,epsfig}
\usepackage[noadjust]{cite}
\usepackage{mcite}
\usepackage{amsfonts,helvet}
\usepackage{fancyhdr}
\usepackage{threeparttable}
\usepackage{epsf,epsfig}
\usepackage{amsthm}
\usepackage{amsmath}
\usepackage{siunitx}
\usepackage{amssymb}
\usepackage{xcolor}

\usepackage[colorlinks=true, linkcolor=blue]{hyperref}

\usepackage{dsfont}
\usepackage{subfigure}
\usepackage{color}
\usepackage{enumerate}
\usepackage{gensymb}
\usepackage{cancel}
\usepackage{lipsum}
\usepackage{mathtools}
\usepackage{cuted}
\usepackage{bbm}
\usepackage[linesnumbered,ruled]{algorithm2e}

\newtheorem{theorem}{Theorem}

\newtheorem{corollary}{Corollary}

\newtheorem{remark}{Remark}

\usepackage{eucal}

\SetKw{KwBy}{by}

\begin{document}

\title{Efficient RF Chain Selection for MIMO Integrated Sensing and Communications: A Greedy Approach}

\author{Subin~Shin, Seongkyu~Jung, Jinseok~Choi, and Jeonghun~Park

\thanks{
This work was supported by Institute of Information \& communications Technology Planning \& Evaluation (IITP) grant funded by the Korea government(MSIT) (No. RS-2024-00395824, Development of Cloud virtualized RAN (vRAN) system supporting upper-midband), in part by Institute of Information \& communications Technology Planning \& Evaluation (IITP) grant funded by the Korea government(MSIT) (No. RS-2024-00397216, Development of the Upper-mid Band Extreme massive MIMO(E-MIMO)).
S. Shin, S. Jung and J. Park are with the School of Electrical and Electronic Engineering, Yonsei University, South Korea (e-mail:{\texttt{sbshin@yonsei.ac.kr, wjdtjd963@yonsei.ac.kr, jhpark@yonsei.ac.kr}}.)  
J. Choi is with the School of Electrical Engineering, Korea Advanced Institute of Science and Technology (KAIST), South Korea (e-mail:{\texttt{jinseok@kaist.ac.kr}}).
}  
} 
 
\maketitle \setcounter{page}{1} 
\begin{abstract} 
In multiple-input multiple-output integrated sensing and communication (MIMO ISAC) systems, radio frequency chain (i.e., RF chain) selection plays a vital role in reducing hardware cost, power consumption, and computational complexity. 
However, designing an effective RF chain selection strategy is challenging due to the disparity in performance metrics between communication and sensing—mutual information (MI) versus beam-pattern mean-squared error (MSE) or the Cramér-Rao lower bound (CRLB). 
To overcome this, we propose a low-complexity greedy RF chain selection framework maximizing a unified MI-based performance metric applicable to both functions. 
By decomposing the total MI into individual contributions of each RF chain, we introduce two approaches: greedy eigen-based selection (GES) and greedy cofactor-based selection (GCS), which iteratively identify and remove the RF chains with the lowest contribution.
We further extend our framework to beam selection for beamspace MIMO ISAC systems, introducing diagonal beam selection (DBS) as a simplified solution. 
Simulation results show that our proposed methods achieve near-optimal performance with significantly lower complexity than exhaustive search, demonstrating their practical effectiveness for MIMO ISAC systems.
\end{abstract}

\section{Introduction} \label{sec_intro}

Integrating sensing capabilities into wireless communication systems, dubbed as integrated sensing and communication (ISAC), has been recognized as a key technology for 6G cellular networks \cite{liu:jsac:22}. 
With ISAC, it is anticipated that cellular networks enhance situational awareness, enabling more intelligent versatile communication strategies. 
In particular, thanks to abundant spatial degrees-of-freedom (DoF), multiple-input multiple-output (MIMO) systems facilitates both communication and sensing (i.e., MIMO radar and MIMO communications), fostering extensive research on MIMO ISAC \cite{zhuo:commmag:24}. 

In the context of MIMO ISAC, one of the most active research areas is transmit beamforming optimization, leveraging the well-established foundations of beamforming techniques developed for MIMO communications.
For instance, in \cite{liu:twc:18, liu:tsp:18}, a semidefinite program (SDP)–based beamforming optimization method was developed under the assumption that communication and sensing share the same hardware (antenna) and spectrum. 
In \cite{liu:tsp:20, hua:tvt:23}, extending \cite{liu:twc:18, liu:tsp:18}, designated sensing symbols were used, wherein a joint beamforming optimization approach was proposed to minimize the mean-squared error (MSE) of the sensing beam-pattern. 
In \cite{choi:twc:24, kim:arxiv:24, kimWioptIsac}, also considering the sensing beam-pattern MSE as the sensing performance metric as in \cite{liu:twc:18, liu:tsp:18, liu:tsp:20, hua:tvt:23}, 
the impact of imperfect channel state information (CSI) acquisition error was taken into account in the communication performance; so as to enable the joint design of ISAC beamforming strategies.
Distinct from the the sensing beam-pattern MSE mainly considered in the aforementioned studies, in \cite{liu:tsp:22}, the Cramér-Rao lower bound (CRLB) was used as the sensing performance metric and a joint beamforming optimization method was proposed that minimizes the CRLB with given signal-to-interference-plus-noise ratio (SINR) constraints. 
Also, adopting the CRLB as the sensing performance metric, \cite{xiong:tit:23, hua:twc:24} revealed the fundamental trade-off between mutual information (MI) and the CRLB (or the Fisher Information Matrix, FIM) in MIMO ISAC systems. 


In addition to beamforming optimization, another critical problem for MIMO ISAC is the radio frequency (RF) chain\footnote{
Throughout this paper, the term RF chain is used, for simplicity, to collectively denote the hardware path and its connected antenna(s), following the convention in antenna/RF chain selection literature. 
For consistency, we refer to antenna selection techniques in prior work as RF chain selection. 
} selection. 
As the number of RF chains increases, hardware cost, power consumption, and computational complexity also grow significantly. This makes it essential to identify a subset of RF chains that can achieve near-optimal performance \cite{sanay:commmag:04, amadori:tcom:15}. Motivated by this, extensive research has been conducted on RF chain selection, primarily in the context of either sensing or communication, considered separately. 


In the context of sensing, several optimization techniques have been explored for addressing a sparse array beamforming problem, such as the alternating direction method of multipliers (ADMM) \cite{huang:tsp:23}, the quadratically constrained quadratic program (QCQP) \cite{hamza:tsp:19}, and the regularization based convex relaxation \cite{wang:tsp:21}. 
Beyond the optimization based approaches, \cite{arash:twc:23} developed a greedy approach that sequentially finds RF chains to minimize the CRLB, offering a computationally efficient RF chain selection strategy.

In the communication domain, there are also various RF chain selection methods in the literature. 
In \cite{gao:tcom:15}, a convex relaxation method was proposed for transmit RF chain selection by relaxing a binary problem. 
In \cite{silva:tcom:21}, a joint approach that combines block coordinate descent, semidefinite relaxation, and weighted minimum mean-squared error (MMSE) was developed to handle binary selection variables. 
However, these convex relaxation-based methods typically require to use off-the-shelf optimization tools like CVX, leading to extremely high computational complexity \cite{jh:twc:23rsma}. 
To reduce computational complexity, in \cite{ghravi:tsp:04, lim:tvt:09, lin:tvt:12, amadori:tcom:15}, sequential RF chain selection techniques were proposed. 
In these approaches, the sum capacity is decomposed into multiple components, each corresponding to the contribution of the RF chain to the overall capacity. 
Correspondingly, this decomposition enables a greedy selection process with significantly reduced computational burden. 
Building upon the decomposition technique developed in \cite{ghravi:tsp:04, lim:tvt:09, lin:tvt:12, amadori:tcom:15}, a joint precoding and beam selection strategy was investigated in \cite{gao:jsac:16} for a beamspace MIMO system. 
In \cite{choi:tcom:20, parkADCtcom}, the decomposition framework was extended to incorporate the effect of low-resolution quantizers, allowing the RF chain selection process to account for quantization-induced distortion.

Despite its importance, however, RF chain selection in MIMO ISAC systems has received relatively limited attention. 
In \cite{caire:icc:24}, a dynamic programming-based optimization framework was used to identify candidate RF chain subsets; thereafter the transmit covariance matrix was optimized based on the selected RF chain subset. 
In \cite{liu:globecom:23}, the ADMM combined with majorization-minimization (MM) was applied to jointly optimize the ISAC beamforming matrix and the RF chain selection vector.
Unfortunately, existing RF chain selection methods rely on computationally intensive optimization solvers such as CVX, which limits their practical applicability and scalability. This motivates a low-complexity RF chain selection for MIMO ISAC systems.

A critical challenge for devising low-complexity RF chain selection for MIMO ISAC lies in the fundamental disparity between sensing and communication performance metrics (e.g., the beam-pattern MSE or CRLB for sensing vs. the MI for communication), which significantly complicates the design process. 
Interestingly, the seminal work \cite{bell:tit:93} introduced the idea of using MI as an information-theoretic sensing performance metric. This concept was later extended and adopted in various studies on sensing waveform design \cite{yang:taes:07, tang:tsp:10, tang:tsp:19, sun:tgrs:21}.
More recently, motivated by the effectiveness of MI as a unified performance measure, the MI-based sensing metric, often referred to as sensing MI, has been widely adopted in MIMO ISAC scenarios \cite{ouyang:wcl:23, peng:wcl:24, wang:tcom:24}. 
However, to the best of our knowledge, RF chain selection for MIMO ISAC systems has not yet been investigated in the context of unified MI-based performance metrics.

In this paper, we propose low-complexity RF chain selection methods for MIMO ISAC systems, by harnessing a unified MI-based performance metric \cite{ouyang:commmag:23}. 
To this end, we first characterize communication MI and sensing MI, and formulate RF chain subset selection problems for both the transmit and receive sides, whose objective is to maximize the weighted sum of normalized MI. 
However, the optimization problem remains NP-hard, leading to prohibitively high computational complexity. 
To address this challenge, leveraging the fact that both sensing and communication performance metrics are expressed in terms of MI, we transform the original RF chain subset selection problem into a sequential RF chain selection problem. 
To solve this, we devise two different MI decomposition-based greedy selection algorithms: greedy eigen-based selection (GES), which leverages eigenvalue decomposition (EVD) and greedy cofactor-based selection (GCS), which is based on matrix cofactor computations. 
Both methods decompose the overall MI into individual RF chain contributions and 
iteratively identify the RF chain subset by sequentially eliminating the RF chain with the lowest contribution in a greedy manner. 

Further, we also show that the proposed GES and GCS are not limited to RF chain selection in fully digital architectures, but are also directly applicable to beam selection in beamspace MIMO ISAC systems with hybrid analog-digital architectures, highlighting the versatility of the proposed methods. 
In addition, under an asymptotic regime where the number of antennas becomes very large, we show that GCS further reduces to a simple diagonal beam selection (DBS), offering an alternative with even lower computational complexity. 

Through extensive numerical results, we demonstrate that the proposed methods outperform baseline approaches. In particular, we demonstrate that the proposed algorithms consistently achieve near-optimal performance across a wide range of average signal-to-noise ratio (SNR) values and the number of active RF chains. 
Notably, the proposed methods also yield considerable energy efficiency gains. 
These gains come from the fact that a carefully selected subset of RF chains can capture most of the ISAC system performance, thereby reducing unnecessary circuit power consumption. 
Finally, by illustrating the Pareto boundary, we demonstrate that the proposed methods achieve a desirable balance between communication and sensing objectives.

The rest of the paper is organized as follows.
Section~\ref{sec_sys} models the considered system and formulates the transmit RF chain selection problem.
Section~\ref{sec_prob_sol} proposes GES and GCS to solve the formulated problem and demonstrates their applicability to the receive RF chain selection problem.
Section~\ref{sec_beam} illustrates how our proposed methods can be readily applied in the beamspace domain and demonstrates the reduction of GCS to simplified DBS in asymptotic cases.
Section~\ref{sec_simul} validates the performance of the proposed methods through extensive numerical results.
Section~\ref{sec_concl} concludes the paper. 

\textbf{Notations}: 
Boldface uppercase and lowercase letters denote matrices and vectors (e.g., $\mathbf{A}$, $\mathbf{a}$), respectively. 
$[\mathbf{A}]_{i,:}$, $[\mathbf{A}]_{:,j}$, and $[\mathbf{A}]_{i,j}$ denote the $i$-th row, $j$-th column, and $(i,j)$-th entry of $\mathbf{A}$, while $\mathbf{A}_{ij}$ denotes a submatrix.
$\mathbf{I}_N$ denotes the $N \times N$ identity matrix.
Superscripts $(\cdot)^{\sf H}$, $(\cdot)^{\sf T}$, $(\cdot)^{-1}$ denote Hermitian, transpose, and inversion; $\mathbf{A}^{-{\sf H}}=(\mathbf{A}^{\sf H})^{-1}$.
$\CMcal{A} = \{a_1, \ldots, a_n\}$ denotes a set, and $\CMcal{A}^c$ its complement.
$|\cdot|$ denotes determinant for matrices and cardinality for sets.
The operator $\operatorname{diag}(\cdot)$ creates a diagonal matrix from a vector or extracts diagonal entries from a matrix.
$\mathbb{C}$, $\mathbb{R}$ and $\mathbb{B}$ denote the complex numbers, real numbers and binary field, respectively.

\section{System Model} \label{sec_sys}
\begin{figure}[t]     
\centerline{\resizebox{0.80\columnwidth}{!}{\includegraphics{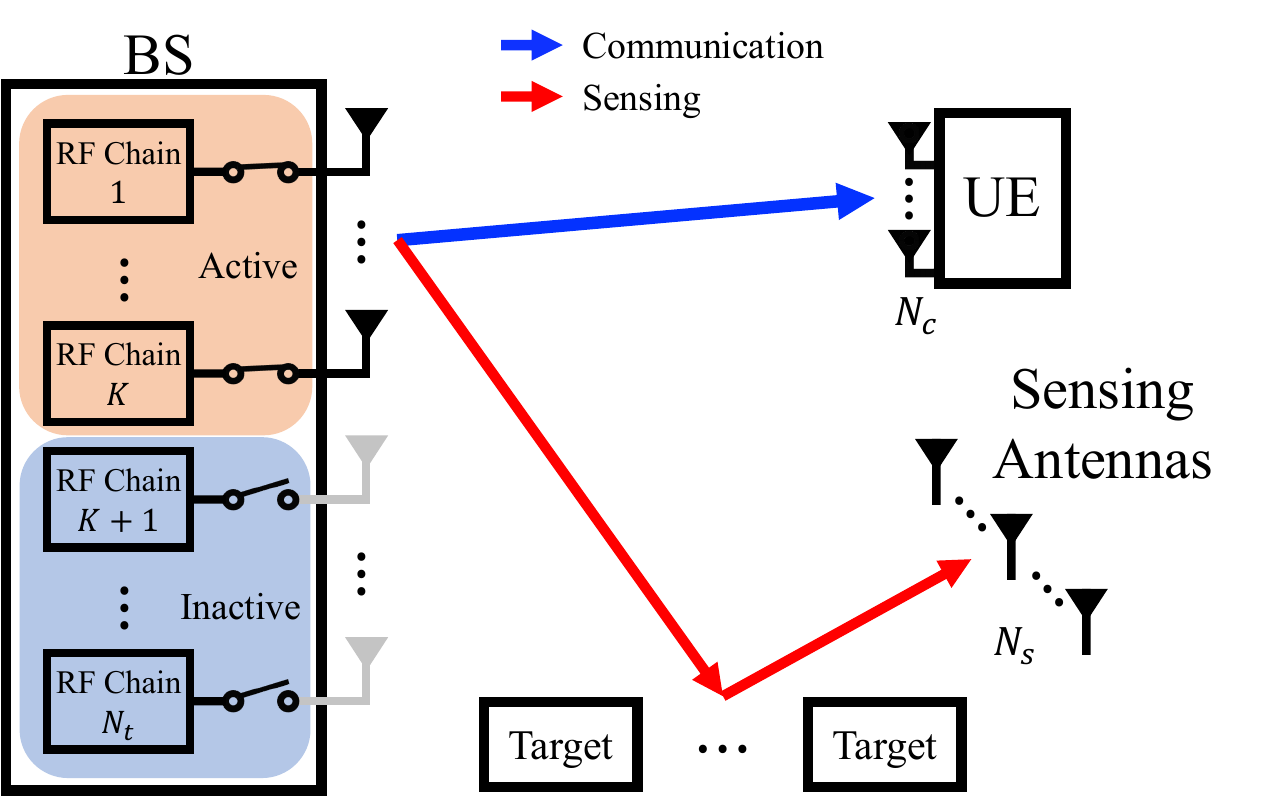}}}
     \caption{
     The considered bistatic MIMO ISAC system with $K$ selected RF chains.
     }\label{fig:sys}
\end{figure}

We consider the MIMO ISAC system in which a base station (BS) with $N_t$ antennas simultaneously communicates with a user equipment (UE) equipped with $N_c$ antennas and performs target sensing in a bistatic setup, where the $N_s$ sensing antennas are located separately from the BS.
\footnote{In the bistatic setup, the BS and sensing receiver are linked by a dedicated backhaul (e.g., optical fiber), so the transmit signal is known at the sensing receiver as a reference waveform \cite{xiong:tit:23}.}
In all cases, we consider a uniform linear array (ULA) configuration, where each antenna is connected to a dedicated RF chain, so that selecting RF chains directly corresponds to antenna selection; Fig.~\ref{fig:sys} illustrates the considered system.

\subsection{Channel and Target Response Model}
The communication channel matrix from the BS to the UE is given as
\begin{align}
    \mathbf{H}_c = \sum_{\ell =1}^{L_{p}} a_{c, \ell} \mathbf{r}(\phi_{\ell}^c) \mathbf{t}^{\sf H}(\theta_{\ell}^c) \in \mathbb{C}^{N_c \times N_t} \label{eq:full_commch},
\end{align} where
$L_{p}$ is the number of paths, 
$a_{c, \ell} \in \mathbb{C}$ is the attenuation coefficient of $\ell$-th path drawn from $\mathcal{CN}(0, 1)$.
$\mathbf{t}(\theta_{\ell}^c)\in \mathbb{C}^{N_t}$ and $\mathbf{r}(\phi_{\ell}^c)\in \mathbb{C}^{N_c}$ are transmit and receive array response vector, respectively.
Under half-wavelength spacing, $\mathbf{t}(\theta_{\ell}^c)$ and $\mathbf{r}(\phi_{\ell}^c)$ are given as
\begin{align}
    \mathbf{t}(\theta_{\ell}^c) &= [1, e^{-j\pi \sin{\theta_{\ell}^c}},\ldots, e^{-j\pi(N_t-1)\sin{\theta_{\ell}^c}}]^{\sf T} , \\
    \mathbf{r}(\phi_{\ell}^c) &= [1, e^{-j\pi \sin{\phi_{\ell}^c}},\ldots, e^{-j\pi(N_c-1)\sin{\phi_{\ell}^c}}]^{\sf T} ,
\end{align} where
$\theta_{\ell}^c$ and $\phi_{\ell}^c$ are the angle of departure (AoD) and angle of arrival (AoA) of $\ell$-th path, respectively.

Now we model the sensing target response matrix (TRM).
Following \cite{tang:tsp:19}, we consider widely separated sensing antennas and model the target response from the $\ell$-th target as $[\mathbf{\bar r}_{\ell}]_{i,:} = (d_{i\ell})^{-\rho/2}e^{-j2\pi \frac{d_{i\ell}}{\lambda}}$, where $d_{i\ell}$ is the distance between the $\ell$-th target and the $i$-th sensing antenna, $\lambda$ and $\rho$ are the wavelength of carrier frequency and the path-loss exponent, and $(d_{i\ell})^{-\rho/2}$ captures the path-loss induced by $d_{i\ell}$.
Using $\mathbf{\bar r}_\ell$, we express the TRM from the BS to the sensing antennas as follows:
\begin{align}
    \mathbf{H}_s = \sum_{\ell =1}^{L_{t}} a_{s, \ell} \mathbf{\bar r}_{\ell} \mathbf{t}^{\sf H}(\theta_{\ell}^s) \in \mathbb{C}^{N_s \times N_t}, \label{eq:full_trm}
\end{align} where
$L_{t}$ is the number of targets and $a_{s,\ell} \in \mathbb{C} \sim \mathcal{CN}(0,(\sigma_{\ell}^s)^2)$ is the complex reflection coefficient. 
We note that this follows Swerling-I model \cite{swerling:ire:1960}. 

Equations \eqref{eq:full_commch} and \eqref{eq:full_trm} assume that all transmit RF chains are used.
To represent the use of a subset of size $K$, we define the index set of active transmit RF chains as $\CMcal{N}_t = \{n_1, \ldots, n_K \}$ with $|\CMcal{N}_t| = K$ and the corresponding selection matrix as
\begin{align}
    \mathbf{S}(\CMcal{N}_t) = [\mathbf{e}_{n_1}, \mathbf{e}_{n_2},\ldots, \mathbf{e}_{n_K}] \in \mathbb{B}^{N_t \times K},
\end{align} 
where $\mathbf{e}_n \in \mathbb{B}^{N_t}$ is the $n$-th standard basis vector in which the $n$-th entry is one and all other entries are zeros. 
Using the selection matrix $\mathbf{S}(\CMcal{N}_t)$, we denote the channel and TRM corresponding to $\CMcal{N}_t$ as follows:
\begin{align}
    \mathbf{H}_c(\CMcal{N}_t) &= \mathbf{H}_c \mathbf{S}(\CMcal{N}_t) \in \mathbb{C}^{N_c \times K}, \\
    \mathbf{H}_s(\CMcal{N}_t) &= \mathbf{H}_s \mathbf{S}(\CMcal{N}_t) \in \mathbb{C}^{N_s \times K}.
\end{align}

\subsection{Signal Model}
Following \cite{stoica:tsp:07}, the transmit signal at time slot $t$ is $\mathbf{x}(t) = [x_1(t), \ldots, x_{N_t}(t)]^{\sf T}\in \mathbb{C}^{N_t}$ with $ \mathbf{x}(t) \sim \mathcal{CN}(\mathbf{0}, P\mathbf{I}_{N_t})$ for the average transmit power $P$, and that over $T$ time slots is $\mathbf{X} = [\mathbf{x}(1), \ldots, \mathbf{x}(T)] \in \mathbb{C}^{N_t \times T}$.
Considering transmit RF chain selection, we express the transmit signal at time slot $t$ and over $T$ time slot as
\begin{align}
    \mathbf{x}(t;\CMcal{N}_t) &= \mathbf{S}^{\sf T}(\CMcal{N}_t)\mathbf{x}(t) \in\mathbb{C}^{K}, \\
    \mathbf{X}(\CMcal{N}_t) &= \mathbf{S}^{\sf T}(\CMcal{N}_t) \mathbf{X} 
    = \big[\mathbf{x}(1;\CMcal{N}_t),\ldots, \mathbf{x}(T;\CMcal{N}_t) \big] \in\mathbb{C}^{K\times T},
\end{align} respectively.
Subsequently, the received signal at the communication UE during $T$ time slots is written as 
\begin{align}
    \mathbf{Y}_c(\CMcal{N}_t) = \mathbf{H}_c(\CMcal{N}_t) \mathbf{X}(\CMcal{N}_t) + \mathbf{Z}_c \in \mathbb{C}^{N_c \times T} \label{eq:comm_signal},
\end{align} where 
$\mathbf{Z}_c \in \mathbb{C}^{N_c\times T}$ is the noise whose columns are independently and identically distributed (i.i.d.) as $\mathcal{CN}(\mathbf{0}, \sigma_c^2 \mathbf{I}_{N_c})$, with $\sigma_c^2$ the communication noise variance.

Similarly, the received signal at sensing antennas during $T$ time slots is 
\begin{align}
    \mathbf{Y}_s^{\sf H}(\CMcal{N}_t) = \mathbf{X}^{\sf H}(\CMcal{N}_t) \mathbf{H}_s^{\sf H}(\CMcal{N}_t)  + \mathbf{Z}_s^{\sf H} \in \mathbb{C}^{T \times N_s}  \label{eq:sensing_signal},
\end{align} where
$\mathbf{Z}_s^{\sf H} \in \mathbb{C}^{T \times N_s}$ is the noise whose columns are i.i.d. $\mathcal{CN}(\mathbf{0}, \sigma_{s}^2\mathbf{I}_T)$, with $\sigma_s^2$ the sensing noise variance.
For notational simplicity, the two noise variances are assumed identical and denoted by $\sigma^2$, and we denote the average transmit power-to-noise ratio $\frac{P}{\sigma^2}$ by $\gamma$.

\subsection{Performance Metrics and Problem Formulation}
In the communication system, the MI between the received signal ${\bf{Y}}_c(\CMcal{N}_t)$ and the transmit signal ${\bf{X}}(\CMcal{N}_t)$ conditioned on the channel ${\bf{H}}_c(\CMcal{N}_t)$ measures the achievable rate \cite{xie:tcom:25}. Based on \eqref{eq:comm_signal}, the communication MI is
\begin{align}
& I_c(\CMcal{N}_t)=  I({\bf{X}}(\CMcal{N}_t); {\bf{Y}}_c(\CMcal{N}_t) | {\bf{H}}_c(\CMcal{N}_t))  \nonumber \\
      &= T \log_2{\left| \mathbf{I}_{K} + \gamma \mathbf{H}_c^{\sf H}(\CMcal{N}_t)\mathbf{H}_c(\CMcal{N}_t)\right|}, \label{eq:before_CMI}\\
     &= T \log_2{\left| \mathbf{I}_{N_c} + \gamma \mathbf{H}_c(\CMcal{N}_t) \mathbf{H}_c^{\sf H}(\CMcal{N}_t)\right|}, \label{eq:CMI}
\end{align} where
\eqref{eq:CMI} is the result of applying the Weinstein-Aronszajn identity $|\mathbf{I}+\mathbf{A}\mathbf{B}| = |\mathbf{I}+\mathbf{B}\mathbf{A}|$.

For the sensing system, we adopt the sensing MI as the sensing performance metric, defined as the MI between the received signal ${\bf{Y}}_s(\CMcal{N}_t)$ and the TRM ${\bf{H}}_s(\CMcal{N}_t)$ conditioned on the transmit signal ${\bf{X}}(\CMcal{N}_t)$.
As shown in \cite{wang:tcom:24,xiong:tit:23}, when the number of time slots $T$ is sufficiently large, the sample covariance matrix of the transmit signal is approximated by its statistical covariance as
\begin{align}
    \frac{1}{T}\mathbf{X}(\CMcal{N}_t)
    \mathbf{X}^{\sf H}(\CMcal{N}_t)
    \approx
    \mathbb{E}\left[
    \mathbf{x}(t;\CMcal{N}_t)
    \mathbf{x}^{\sf H}(t;\CMcal{N}_t)
    \right] 
    = P\mathbf{I}_K.
\end{align}
Assuming that the columns of $\mathbf{H}_s^{\sf H}(\CMcal{N}_t)$ are independent, as in \cite{tang:tsp:19}, the sensing MI based on \eqref{eq:sensing_signal} is given by
\begin{align}
    I_s (\CMcal{N}_t)
    &= \sum_{n=1}^{N_s} \log_2{\left| \mathbf{I}_{K} + \frac{1}{\sigma^2} \mathbf{R}_{{\sf T}, n}(\CMcal{N}_t) \mathbf{X}(\CMcal{N}_t) \mathbf{X}^{\sf H}(\CMcal{N}_t) \right|}, \\
    &\approx \sum_{n=1}^{N_s} \log_2{\left| \mathbf{I}_{K} + \gamma T \mathbf{R}_{{\sf T}, n}(\CMcal{N}_t) \right|}, \label{eq:SMI_each}
\end{align}
where $\mathbf{R}_{{\sf T}, n}(\CMcal{N}_t)$ is the covariance matrix of the $n$-th column vector of $\mathbf{H}_s^{\sf H}(\CMcal{N}_t)$, defined as
\begin{align}
    & \mathbf{R}_{{\sf T}, n}(\CMcal{N}_t) = \sum_{\ell=1}^{L_{t}} \kappa_{n \ell} \mathbf{S}^{\sf T}(\CMcal{N}_t)\mathbf{t}(\theta_\ell)\mathbf{t}^{\sf H}(\theta_\ell)\mathbf{S}(\CMcal{N}_t) \in \mathbb{C}^{K \times K}.  \label{eq:sen_cov}
\end{align}
In \eqref{eq:sen_cov}, $\kappa_{n\ell} = (\sigma_\ell^s)^2\left|\left[ \mathbf{\bar r}_\ell \right]_{n,:}\right|^2$ captures the reflection of the $\ell$-th target and the path-loss between it and the $n$-th sensing antenna.

Based on \eqref{eq:CMI} and \eqref{eq:SMI_each}, 
we formulate the transmit RF chain selection problem that maximizes the weighted sum of normalized MI as follows: 
\begin{subequations}
    \begin{align}
    \underset{\CMcal{N}_t}{\max} \quad & \frac{\omega_c}{T}I_c(\CMcal{N}_t) + \frac{\omega_s}{N_s}I_s(\CMcal{N}_t) \\
    \textrm{s.t.}\quad & |\CMcal{N}_t| = K,
\end{align} \label{eq:ori_prob}
\end{subequations} 
where $0\leq \omega_c,\omega_s\leq1$ are weights to balance the trade-off between the communication MI and the sensing MI with $\omega_c + \omega_s = 1$. 
In \eqref{eq:ori_prob}, note that the normalized communication MI and sensing MI are used for tractable and balanced optimization.
Unfortunately, \eqref{eq:ori_prob} is an NP-hard problem that incurs prohibitively high computational complexity.
In the next section, we propose greedy RF chain selection methods that address \eqref{eq:ori_prob} while significantly reducing computational complexity, and, since \eqref{eq:ori_prob} only considers the transmit side, we further investigate the receive RF chain selection problem in Section \ref{subsec_rxsel}.

\begin{remark}[Justification of the sensing MI as a sensing metric]
\label{rem:smi}
\normalfont
Although less conventional than the CRLB or beam-pattern MSE,
the sensing MI is not an ad hoc metric. The exact sensing MI
prior to the large-$T$ approximation in \eqref{eq:SMI_each},
$I_s=I(\mathbf{Y}_s;\mathbf{H}_s\mid\mathbf{X})$, quantifies
the information about the TRM contained in the received signal
for a given transmit signal. If a target-parameter vector
$\boldsymbol{\xi}$ fully specifies the TRM through a
deterministic mapping
$\mathbf{H}_s=\mathbf{H}_s(\boldsymbol{\xi})$, then
$\boldsymbol{\xi}\to\mathbf{H}_s\to\mathbf{Y}_s$ gives
$I(\mathbf{Y}_s;\boldsymbol{\xi}\mid\mathbf{X})
=I(\mathbf{Y}_s;\mathbf{H}_s\mid\mathbf{X})$
\cite{liu:isit:23}.

This interpretation is connected to classical sensing criteria
in two complementary senses.
(i) Distortion limit \cite{liu:isit:23}: for any estimate
$\hat{\boldsymbol{\xi}}$ formed from $\mathbf{Y}_s$ given
$\mathbf{X}$, the rate--distortion theorem yields
$\mathbb{E}[d(\boldsymbol{\xi},\hat{\boldsymbol{\xi}})]
\geq D(I(\boldsymbol{\xi};\mathbf{Y}_s\mid\mathbf{X}))$,
where $D(\cdot)$ is the non-increasing distortion--rate
function. Squared-error distortion gives an MSE-type lower
bound. For binary target-presence detection, the Hamming
distortion is related to the detection and false-alarm
probabilities, and, under a fixed false-alarm probability,
a smaller distortion implies a larger detection probability.
This is a fundamental converse bound and is not generally
guaranteed to be achieved by an arbitrary estimator or detector.
(ii) Gaussian TRM relation \cite{xie:tcom:25}: for a Gaussian
$\mathbf{h}_s=\mathrm{vec}(\mathbf{H}_s^{\sf H})
\sim\mathcal{CN}(\mathbf{0},\mathbf{R})$, when $\mathbf{R}$
is nonsingular, Proposition~4 of \cite{xie:tcom:25} gives
$\boldsymbol{\Psi}_{\mathrm{BCRLB}}(\mathbf{X})
=\boldsymbol{\Psi}_{\mathrm{MMSE}}(\mathbf{X})
=\boldsymbol{\Psi}_{\mathrm{LMMSE}}(\mathbf{X})$ and
$I_s=\log_2|\mathbf{R}|
-\mathbb{E}_{\mathbf{X}}
[\log_2|\boldsymbol{\Psi}_{\mathrm{BCRLB}}(\mathbf{X})|]$.
Thus, for a fixed $\mathbf{R}$, maximizing the exact sensing
MI is equivalent to minimizing the expected log-determinant
of the BCRLB matrix. The exact equivalence is therefore with
the log-determinant BCRLB criterion, while trace-based
EMMSE/EBCRLB criteria remain related but are not identical
objectives. The sensing MI has also been adopted in recent
MIMO ISAC studies
\cite{ouyang:wcl:23,peng:wcl:24,wang:tcom:24}, and its
large-$T$ approximation in \eqref{eq:SMI_each} enables the
unified objective in \eqref{eq:ori_prob}.
\end{remark}

\section{Proposed RF Chain Selection Methods}\label{sec_prob_sol}


To avoid the high computational complexity arising from evaluating all possible transmit RF chain combinations, we transform \eqref{eq:ori_prob} into a sequential transmit RF chain selection problem: 
\begin{align}
    \underset{j \in \CMcal{\bar N}_t }{\max} \quad & \frac{\omega_c}{T}I_c(\CMcal{\bar N}_t \setminus \{j\}) + \frac{\omega_s}{N_s}I_s(\CMcal{\bar N}_t\setminus \{j\}), \label{eq:seq_prob}  
\end{align} 
where \(\bar{\CMcal{N}}_t\) is the index set of remaining transmit RF chains during the sequential selection process.
The problem \eqref{eq:seq_prob} greedily chooses the index \(j \in \bar{\CMcal{N}}_t\) whose removal incurs the minimal loss in the weighted sum of the normalized communication and sensing MI.
After finding the optimum $j^*$ by solving \eqref{eq:seq_prob}, the remaining RF chain set $\bar{\CMcal{N}}_t$ is updated as $\bar{\CMcal{N}}_t \leftarrow \bar{\CMcal{N}}_t \setminus \{j^*\}$ and the sequential selection process is repeated until $|\bar{\CMcal{N}}_t| = K$.


A key step of running the greedy method for the problem \eqref{eq:seq_prob} is sequentially finding the RF chain that has the lowest contribution to the objective function; to this end, we present two different MI decomposition techniques, each corresponding to a distinct RF chain selection method: 1) GES and 2) GCS.


\subsection{Greedy Eigen-based Selection}

The GES method leverages the EVD to decompose the MI expressions, whose main result is presented in Theorem~\ref{eq_theo_ges}.
{\theorem \label{eq_theo_ges}
Let $j$ be an index of the RF chain included in $\CMcal{\bar N}_t$ and
$\mathbf{R}_{{\sf T}, n} = \mathbf{\tilde P}_n\mathbf{\tilde \Lambda}_n\mathbf{\tilde P}_n^{\sf H}$ be the decomposition of eigenvalues.
Then, communication MI and sensing MI are decomposed as follows:
\begin{align}
    I_c(\CMcal{\bar N}_t \setminus \{j\}) &= I_c(\CMcal{\bar N}_t) - T \log_2{\left( (1- \gamma \alpha_j)^{-1} \right)},\\
    I_s(\CMcal{\bar N}_t \setminus \{j\}) &= I_s(\CMcal{\bar N}_t) - \sum_{n =1}^{N_s}\log_2{\left( (1 -\gamma T\beta_{n,j})^{-1} \right)}, 
\end{align}
where
\begin{align}
    \alpha_j &= \mathbf{h}_j^{\sf H} \mathbf{A} \mathbf{h}_j,\; \mathbf{h}_j = [\mathbf{H}_c(\bar{\CMcal{N}}_t)]_{:, j}, \label{eq:alp} \\
    \mathbf{A} &= \left( \mathbf{I}_{N_c} +\gamma \mathbf{H}_c(\CMcal{\bar N}_t) \mathbf{H}_c^{\sf H}(\CMcal{\bar N}_t) \right)^{-1}, \label{eq:A} \\
    \beta_{n,j}&= \mathbf{g}_{n, j}^{\sf H} \mathbf{B}_n \mathbf{g}_{n,j}, \; \mathbf{g}_{n,j} = [\mathbf{G}_n^{\sf H}(\bar{\CMcal{N}}_t)]_{:, j},\label{eq:beta}\\
    \mathbf{B}_n &= \left(  \mathbf{I}_{r_n} + \gamma T
    \mathbf{G}_{n}^{\sf H}(\CMcal{\bar N}_t) \mathbf{G}_{n}(\CMcal{\bar N}_t) \right)^{-1}, \label{eq:B} \\
    \mathbf{G}_n(\CMcal{\bar N}_t) &= \mathbf{S}^{\sf T}(\CMcal{\bar N}_t)\mathbf{P}_n\mathbf{\Lambda}_n^{1/2},
\end{align}
where $r_n$ is the rank of $\mathbf{R}_{{\sf T},n}$ and $\mathbf{P}_n \in \mathbb{C}^{N_t \times r_n}$ is the submatrix of $\mathbf{\tilde P}_n \in \mathbb{C}^{N_t \times N_t}$, while the diagonal entries of $\mathbf{\Lambda}_n \in \mathbb{R}^{r_n \times r_n}$ are non-zero eigenvalues of $\mathbf{R}_{{\sf T}, n}$.
\begin{proof}
    See Appendix~\ref{appA}.
\end{proof}
}

By Theorem~\ref{eq_theo_ges}, we rewrite \eqref{eq:seq_prob} as 
\begin{align}
    &\frac{\omega_c}{T}I_c(\CMcal{\bar N}_t) + \frac{\omega_s}{N_s}I_s(\CMcal{\bar N}_t) \nonumber\\
    &-\underbrace{\log_2{\left( \left( 1-\gamma\alpha_j \right)^{-\omega_c} \prod_{n=1}^{N_s}\left( 1-\gamma T\beta_{n,j} \right)^{-\frac{\omega_s}{N_s}} \right)}}_{\text{The contribution of the $j$-th RF chain}},
\end{align} 
which captures the contribution of the $j$-th RF chain to the weighted sum of normalized MI.
To remove the RF chain with the lowest contribution at each iteration, we reformulate \eqref{eq:seq_prob} as 
\begin{align}
    \underset{j \in \CMcal{\bar N}_t^{(i)} }{\max} \quad &   \left( 1-\gamma\alpha_j^{(i)} \right)^{\omega_c} \prod_{n=1}^{N_s}\left( 1-\gamma T\beta_{n,j}^{(i)} \right)^{\frac{\omega_s}{N_s}}, \label{eq:final_prob1}
\end{align} where $i$ denotes the current iteration number and
$\alpha_j^{(i)}, \beta_{n,j}^{(i)}$ and $\CMcal{\bar N}_t^{(i)}$ are the communication and sensing contribution parameters and the index set of remaining transmit RF chains at the $i$-th iteration, respectively. 

However, updating $\alpha_j^{(i)}$ and $\beta_{n,j}^{(i)}$ in \eqref{eq:alp}, \eqref{eq:beta} requires matrix inversion operations with high computational complexity per iteration, which the following corollary avoids by sequentially updating $\alpha_j^{(i)}, \mathbf{A}^{(i)}$ and $\beta_{n,j}^{(i)}, \mathbf{B}_n^{(i)}$.

{\corollary \label{eq_cor_ges}
Let $J^{(i)}$ denote the index of the RF chain that was removed at the $i$-th iteration.
Then, $\alpha_j^{(i+1)}$ and $\beta_{n,j}^{(i+1)}$ are sequentially updated as 
\begin{align}
    \alpha_j^{(i+1)} &= \alpha_j^{(i)} +|\mathbf{h}_j^{\sf H} \mathbf{a}^{(i)}|^2, \label{eq:up_alp} \\    
    \beta_{n,j}^{(i+1)} &= \beta_{n,j}^{(i)} 
    + |\mathbf{g}_{n,j}^{\sf H} \mathbf{b}_n^{(i)}|^2, 
    \label{eq:up_beta}
\end{align} where
\begin{align}
    \mathbf{a}^{(i)} &= \sqrt{\gamma\left( 1 - \gamma \alpha_{J^{(i)}}^{(i)} \right)^{-1}}
    \mathbf{A}^{(i)} \mathbf{h}_{J^{(i)}} , \label{eq:a}\\
    \mathbf{b}_n^{(i)} &= \sqrt{ \gamma T \left( 1 - \gamma T\beta_{n,{J^{(i)}}}^{(i)} \right)^{-1}}
    \mathbf{B}_n^{(i)} \mathbf{g}_{n,{J^{(i)}}}. \label{eq:b}
\end{align}
For obtaining $\mathbf{a}^{(i+1)}$ and $\mathbf{b}_n^{(i+1)}$,
$\mathbf{A}^{(i+1)}$ and $\mathbf{B}_n^{(i+1)}$ are sequentially updated as
\begin{align}
    \mathbf{A}^{(i+1)} &= \mathbf{A}^{(i)} + \mathbf{a}^{(i)}\mathbf{a}^{(i){\sf H}}, \label{eq:up_A} \\
    \mathbf{B}_n^{(i+1)} &= \mathbf{B}_n^{(i)} + \mathbf{b}_n^{(i)}\mathbf{b}_n^{(i){\sf H}}. \label{eq:up_B}
\end{align}
\begin{proof}
    \text{See Appendix~\ref{appB}.}
\end{proof}}

Using Theorem~\ref{eq_theo_ges} and Corollary~\ref{eq_cor_ges}, we develop the GES algorithm in Algorithm~\ref{alg:GES}.

\begin{algorithm} [t]
\caption{Greedy Eigen-based Selection} \label{alg:GES} 
\textbf{input}: Desired number of RF chains $K$, weighting factors $\omega_c$, $\omega_s$ \\
{\bf{initialize}}: { $i = 0, \CMcal{\bar N}_t^{(0)} = \{1,2,\ldots,N_t\}$ \\
EVD of $\mathbf{R}_{{\sf T}, n}$\\
Compute $\mathbf{A}^{(0)}$, $\mathbf{B}_n^{(0)}$  using \eqref{eq:A}, \eqref{eq:B}\\
Compute $\alpha_j^{(0)}$, $\beta_{n,j}^{(0)}$ using \eqref{eq:alp}, \eqref{eq:beta}
} \\
\While {$|\CMcal{\bar N}_t^{(i)}| > K$}
{
    Select $J^{(i)}$ according to \eqref{eq:final_prob1}
    \\    
    $\CMcal{\bar N}_t^{(i+1)} \leftarrow \CMcal{\bar N}_t^{(i)} \setminus \{J^{(i)}\}$ \\
    \If{$|\CMcal{\bar N}_t^{(i+1)}| = K$}{
    $\CMcal{N}_t \leftarrow \CMcal{\bar N}_t^{(i+1)} $\\
    \Return{$\CMcal{N}_t$}}
    Compute $\mathbf{a}^{(i)}$, $\mathbf{b}_n^{(i)}$ using \eqref{eq:a}, \eqref{eq:b} \\   
    Update $\mathbf{A}^{(i+1)}$, $\mathbf{B}_n^{(i+1)}$ using \eqref{eq:up_A}, \eqref{eq:up_B} \\
    Update $\alpha_j^{(i+1)}$, $\beta_{n,j}^{(i+1)}$ using \eqref{eq:up_alp}, \eqref{eq:up_beta} \\
    $i \leftarrow i+1$
}
\end{algorithm}

\subsection{Greedy Cofactor-based Selection}
Similar to GES, GCS sequentially removes the RF chain with the lowest contribution in a greedy fashion, but characterizes the contribution to the weighted sum of MI by using a cofactor-based inverse matrix representation, as stated in the theorem below.
{\theorem \label{eq_theo_gcs}
Let $j$ be an index of the RF chain included in $\CMcal{\bar N}_t$. 
Then, communication MI and sensing MI are decomposed as follows:
\begin{align}
    I_c(\CMcal{\bar N}_t \setminus \{j\}) &= I_c(\CMcal{\bar N}_t) - T \log_2(\delta_j^{-1}), \\
    I_s(\CMcal{\bar N}_t \setminus \{j\}) &=  I_s(\CMcal{\bar N}_t ) - \sum_{n =1}^{N_s} \log_2{(\varepsilon_{n,j}^{-1})}
\end{align}
where
\begin{align}
    \delta_j &= [\mathbf{D}^{-1}(\CMcal{\bar N}_t)]_{j,j}, \label{eq:delta}\\
    \mathbf{D}(\CMcal{\bar N}_t) &= \mathbf{I}_{|\CMcal{\bar N}_t|} + \gamma \mathbf{H}_c^{\sf H}(\CMcal{\bar N}_t) \mathbf{H}_c(\CMcal{\bar N}_t), \label{eq:D} \\
    \varepsilon_{n,j} &= [\mathbf{E}_n^{-1}(\CMcal{\bar N}_t)]_{j,j}, \label{eq:eps}\\
    \mathbf{E}_n(\CMcal{\bar N}_t) &= \mathbf{I}_{|\CMcal{\bar N}_t|} + \gamma T \mathbf{R}_{{\sf T}, n}(\CMcal{\bar N}_t). \label{eq:E}
\end{align}

\begin{proof}
    See Appendix~\ref{appC}.
\end{proof}
}

By Theorem~\ref{eq_theo_gcs}, \eqref{eq:seq_prob} is reformulated as
\begin{align}
    \underset{j \in \CMcal{\bar N}_t^{(i)} }{\max} \quad &  \left(\delta_j^{(i)}\right)^{\omega_c} \prod_{n=1}^{N_s}\left(\varepsilon_{n,j}^{(i)}\right)^{\frac{\omega_s}{N_s}}, \label{eq:final_prob2}
\end{align} where
$\delta_j^{(i)}, \varepsilon_{n,j}^{(i)}$ are the communication and sensing contribution parameters at the $i$-th iteration.

As in GES, updating the parameters requires matrix inversion operations at each iteration, which the following corollary resolves based on the Schur complement.
{\corollary \label{eq_cor_gcs}
Let $J^{(i)}$ denote the index of the RF chain that was removed at the $i$-th iteration.
Then, 
$\mathbf{D}^{-(i+1)}$ and $\mathbf{E}_n^{-(i+1)}$ are sequentially updated as
\begin{align}
    \mathbf{D}^{-(i+1)} &= \mathbf{\tilde D}_{11}^{-(i)} - \frac{\mathbf{\tilde D}_{12}^{-(i)}\mathbf{\tilde D}_{21}^{-(i)} }{ \mathbf{\tilde D}_{22}^{-(i)} }, \label{eq:up_invD} \\
    \mathbf{E}_n^{-(i+1)}&= \mathbf{\tilde E}_{n11}^{-(i)} - \frac{\mathbf{\tilde E}_{n12}^{-(i)}\mathbf{\tilde E}_{n21}^{-(i)} }{ \mathbf{\tilde E}_{n22}^{-(i)} }, \label{eq:up_invE}
\end{align} where
\begin{align}
    \mathbf{\tilde D}^{-(i)} &= \mathbf{P}^{(i)}\mathbf{D}^{-(i)}\mathbf{P}^{(i) {\sf T}} = 
    \begin{bmatrix}
        \mathbf{\tilde D}_{11}^{-(i)} & \mathbf{\tilde D}_{12}^{-(i)} \\
        \mathbf{\tilde D}_{21}^{-(i)} & \mathbf{\tilde D}_{22}^{-(i)}
    \end{bmatrix} , \\
    \mathbf{\tilde E}_n^{-(i)} &= \mathbf{P}^{(i)}\mathbf{E}_n^{-(i)}\mathbf{P}^{(i){\sf T}} = 
    \begin{bmatrix}
        \mathbf{\tilde E}_{n11}^{-(i)} & \mathbf{\tilde E}_{n12}^{-(i)} \\
        \mathbf{\tilde E}_{n21}^{-(i)} & \mathbf{\tilde E}_{n22}^{-(i)}
    \end{bmatrix}, 
\end{align}
with the permutation matrix,
\begin{align}
     [\mathbf{P}^{(i)}]_{k,:} =
    \begin{cases}
    \mathbf{e}_k^{\sf T}       & \text{for } 1 \leq k < J^{(i)} \\
    \mathbf{e}_{k+1}^{\sf T}   & \text{for } J^{(i)} \leq k < |\CMcal{\bar N}_t^{(i)}| \\
    \mathbf{e}_{J^{(i)}}^{\sf T}       & \text{for } k = |\CMcal{\bar N}_t^{(i)}|
    \end{cases}.
\end{align}
\begin{proof}
    \text{See Appendix~\ref{appD}.}
\end{proof}}

Using Theorem~\ref{eq_theo_gcs} and Corollary~\ref{eq_cor_gcs}, we develop the GCS algorithm in Algorithm~\ref{alg:GCS}. 
\begin{algorithm} [t]
\caption{Greedy Cofactor-based Selection} \label{alg:GCS} 
\textbf{input}: Desired number of RF chains $K$, weighting factors $\omega_c$, $\omega_s$ \\
{\bf{initialize}}: {$i = 0, \CMcal{\bar N}_t^{(0)} = \{1,2,\ldots,N_t\}$ \\
Compute $\mathbf{D}^{-(0)}$, $\mathbf{E}_n^{-(0)}$ using \eqref{eq:D}, \eqref{eq:E} \\
Compute $\delta_j^{(0)}$, $\varepsilon_{n,j}^{(0)}$ using \eqref{eq:delta}, \eqref{eq:eps}
}\\
\While {$|\CMcal{\bar N}_t^{(i)}| > K$}
{
    Select $J^{(i)}$ according to \eqref{eq:final_prob2}
    \\
    $\CMcal{\bar N}_t^{(i+1)} \leftarrow \CMcal{\bar N}_t^{(i)} \setminus \{J^{(i)}\}$ \\
    \If{$| \CMcal{\bar N}_t^{(i+1)} | = K$}{
    $\CMcal{N}_t \leftarrow \CMcal{\bar N}_t^{(i+1)} $\\
    \Return{$\CMcal{N}_t$}}
    
    Update $\mathbf{D}^{-(i+1)}$, $\mathbf{E}_n^{-(i+1)}$ using \eqref{eq:up_invD}, \eqref{eq:up_invE} \\
    Update $\delta_j^{(i+1)}$, $\varepsilon_{n,j}^{(i+1)}$ using \eqref{eq:delta}, \eqref{eq:eps} \\
    $i \leftarrow i+1$
}
\end{algorithm}



\subsection{Comparison of GES and GCS}
\label{subsec_GES_GCS}

Although GES and GCS are derived through different algebraic operations---the EVD-based decomposition in Theorem~\ref{eq_theo_ges} and the cofactor-based decomposition in Theorem~\ref{eq_theo_gcs}---they implement the same greedy RF chain selection rule in exact arithmetic under the same tie-breaking rule.
This is seen by applying the Woodbury matrix identity to $\mathbf{D}(\bar{\CMcal{N}}_t)$ and $\mathbf{E}_n(\bar{\CMcal{N}}_t)$ in \eqref{eq:D} and \eqref{eq:E}:
\begin{align}
\mathbf{D}^{-1}(\bar{\CMcal{N}}_t)
&= \left(\mathbf{I}_{|\bar{\CMcal{N}}_t|}
+\gamma\mathbf{H}_c^{\sf H}(\bar{\CMcal{N}}_t)\mathbf{H}_c(\bar{\CMcal{N}}_t)\right)^{-1} \nonumber \\
&= \mathbf{I}_{|\bar{\CMcal{N}}_t|}
-\gamma\mathbf{H}_c^{\sf H}(\bar{\CMcal{N}}_t)\mathbf{A}\mathbf{H}_c(\bar{\CMcal{N}}_t), \\
\mathbf{E}_n^{-1}(\bar{\CMcal{N}}_t)
&= \left(\mathbf{I}_{|\bar{\CMcal{N}}_t|}
+\gamma T\,\mathbf{G}_n(\bar{\CMcal{N}}_t)\mathbf{G}_n^{\sf H}(\bar{\CMcal{N}}_t)\right)^{-1} \nonumber \\
&= \mathbf{I}_{|\bar{\CMcal{N}}_t|}
-\gamma T\,\mathbf{G}_n(\bar{\CMcal{N}}_t)\mathbf{B}_n\mathbf{G}_n^{\sf H}(\bar{\CMcal{N}}_t),
\end{align}
where $\mathbf{A}$ and $\mathbf{B}_n$ are defined in \eqref{eq:A} and \eqref{eq:B}.
Since $\mathbf{h}_j=\mathbf{H}_c(\bar{\CMcal{N}}_t)\mathbf{e}_j$ and $\mathbf{g}_{n,j}=\mathbf{G}_n^{\sf H}(\bar{\CMcal{N}}_t)\mathbf{e}_j$, the $j$-th diagonal entries reduce to
\begin{align}
\delta_j=[\mathbf{D}^{-1}(\bar{\CMcal{N}}_t)]_{j,j}
&= 1-\gamma\,\mathbf{h}_j^{\sf H}\mathbf{A}\mathbf{h}_j
= 1-\gamma\alpha_j, \\
\varepsilon_{n,j}=[\mathbf{E}_n^{-1}(\bar{\CMcal{N}}_t)]_{j,j}
&= 1-\gamma T\,\mathbf{g}_{n,j}^{\sf H}\mathbf{B}_n\mathbf{g}_{n,j}
= 1-\gamma T\beta_{n,j},
\end{align}
i.e., the GCS contribution factors are exactly the residual factors of the GES selection metric.
Consequently, \eqref{eq:final_prob2} coincides with \eqref{eq:final_prob1} at every iteration, so both methods remove the same RF chain and return the same $\CMcal{N}_t$ with the same weighted sum of MI.
Nevertheless, since they evaluate the contributions in different matrix domains, they incur different computational profiles in both the initialization and the sequential update stages.

In the initialization stage, GES performs the EVD of $\mathbf{R}_{{\sf T},n}$ for all $n$ and forms the compact factors $\mathbf{G}_n=\mathbf{P}_n\boldsymbol{\Lambda}_n^{1/2}$; along with computing $\mathbf{A}^{(0)}$ and $\alpha_j^{(0)}$, and using $\mathbf{G}_n^{\sf H}\mathbf{G}_n=\boldsymbol{\Lambda}_n$ at the full-candidate set, $\mathbf{B}_n^{(0)}$ is diagonal and $\beta_{n,j}^{(0)}$ costs only $\mathcal{O}(r_n)$ per candidate, giving $\mathcal{O}(N_c^3+N_tN_c^2 + N_sN_t^3+N_sN_t r)$, where $r = \max_n r_n$.
GCS instead forms the RF chain index domain inverses $\mathbf{D}^{-(0)}$ and $\mathbf{E}_n^{-(0)}$ and reads off their diagonals, at $\mathcal{O}(N_cN_t^2+(N_s+1)N_t^3)$ without any EVD.

In the sequential update stage, both methods avoid recomputing the inverses from scratch via Corollary~\ref{eq_cor_ges} and Corollary~\ref{eq_cor_gcs}.
At the $i$-th iteration with $m_i=N_t-i$ remaining RF chains, GES applies rank-one updates of $\mathbf{A}$ and $\mathbf{B}_n$ and refreshes all contributions at $\mathcal{O}(N_c^2+N_s r^2+m_i(N_c+N_s r))$, whereas GCS applies the Schur complement down-date at $\mathcal{O}((N_s+1)m_i^2)$.
Summing over the $N_t-K$ removals yields $\mathcal{O}((N_t-K)(N_c^2+N_s r^2)+(N_t^2-K^2)(N_c+N_s r))$ for GES and $\mathcal{O}((N_s+1)(N_t^3-K^3))$ for GCS.
Hence the sequential cost of GES is governed by $r$, while that of GCS is governed by $m_i$.

Accordingly, GES and GCS are preferable in complementary regimes jointly determined by the sensing rank $r$ and the selection depth $N_t-K$, which we confirm through the runtime regime map in Section~\ref{subsec_hmap}.

\subsection{Receive RF chain Selection} \label{subsec_rxsel}



Now we explore the receive RF chain selection, which is simpler than the transmit counterpart because the communication MI and the sensing MI are inherently decoupled at the receiver side; hence, the selection is independently performed at the communication UE and at the sensing antennas without complicated interdependence between the two domains.


Denoting the active communication and sensing receive RF chain sets as $\CMcal{N}_c = \{n_{c, 1}, \ldots, n_{c, K}\}$ and $\CMcal{N}_s= \{n_{s, 1}, \ldots, n_{s, K}\}$, respectively, we rewrite the communication MI \eqref{eq:CMI} and sensing MI \eqref{eq:SMI_each} in terms of $\CMcal{N}_c$ and $\CMcal{N}_s$ as follows:
\begin{align}
     I_c(\CMcal{N}_c) &= T \log_2{\left| \mathbf{I}_{N_t} + \gamma \mathbf{H}_c^{\sf H} (\CMcal{N}_c)\mathbf{H}_c(\CMcal{N}_c)\right|}, \label{eq:CMI_rcvGES} \\
     &= T \log_2{\left| \mathbf{I}_{K} + \gamma \mathbf{H}_c(\CMcal{N}_c)\mathbf{H}_c^{\sf H}(\CMcal{N}_c)\right|}, \label{eq:CMI_rcvGCS} \\
     I_s(\CMcal{N}_s) &= \sum_{n \in \CMcal{N}_s} \log_2{\left| \mathbf{I}_{N_t} + \gamma T \mathbf{R}_{{\sf T}, n} \right|} \label{eq:SMI_rcv},
\end{align} where
\begin{align}
    \mathbf{H}_c(\CMcal{N}_c) &= \mathbf{S}_{r}^{\sf T}(\CMcal{N}_c) \mathbf{H}_c \in \mathbb{C}^{K \times N_t}, \\
    \mathbf{S}_{r}(\CMcal{N}_c) &= [\mathbf{e}_{n_{c,1}}, \mathbf{e}_{n_{c,2}},\ldots, \mathbf{e}_{n_{c, K}}] \in \mathbb{B}^{N_c \times K}.
\end{align}
In \eqref{eq:CMI_rcvGCS} and \eqref{eq:SMI_rcv}, we observe that the communication MI and the sensing MI are affected only by their own respective receive RF chain sets.
Therefore, the sequential receive RF chain selection problem, which aims to maximize the sum of normalized MI, are formulated as two independent optimization problems as follows:
\begin{subequations}
    \begin{align}
    \underset{j \in \CMcal{\bar N}_c }{\max} \quad & \frac{1}{T}I_c(\CMcal{\bar N}_c \setminus \{j\}), 
    \label{eq:prob_comm}\\
    \underset{j \in \CMcal{\bar N}_s }{\max} \quad & \frac{1}{N_s}I_s(\CMcal{\bar N}_s\setminus \{j\}). 
    \label{eq:prob_sen}
\end{align} 
\end{subequations}

For the communication receive RF chain selection \eqref{eq:prob_comm}, \eqref{eq:CMI_rcvGES} and \eqref{eq:CMI_rcvGCS} have the equivalent form as \eqref{eq:CMI} and \eqref{eq:before_CMI}, respectively, in terms of how RF chain removal is represented in the equations.
For this reason, \eqref{eq:CMI_rcvGES} can be applied to Theorem~\ref{eq_theo_ges} and Corollary~\ref{eq_cor_ges}, and \eqref{eq:CMI_rcvGCS} to Theorem~\ref{eq_theo_gcs} and Corollary~\ref{eq_cor_gcs}, so that GES and GCS directly solve \eqref{eq:prob_comm}; the derivation is identical and hence not repeated.

Meanwhile, the sensing RF chain selection \eqref{eq:prob_sen} is even simpler: as shown in \eqref{eq:SMI_rcv}, the contribution of the $n$-th sensing RF chain to the sensing MI is explicitly expressed as $\log_2{\left| \mathbf{I}_{N_t} + \gamma T \mathbf{R}_{{\sf T}, n} \right|}$, so that \eqref{eq:prob_sen} is reformulated as
\begin{align}
    \underset{n \in \CMcal{\bar N}_s }{\min} \quad & \left| \mathbf{I}_{N_t} + \gamma T \mathbf{R}_{{\sf T}, n} \right|, \label{eq:prob_sen_easy}
\end{align}
which is solved by simply comparing $|\mathbf{I}_{N_t} +\gamma T\mathbf{R}_{{\sf T}, n}|$ for all candidate sensing RF chains $n \in \CMcal{N}_s$.

\section{Extension to Beam Selection for Hybrid MIMO ISAC Architecture} \label{sec_beam}
\begin{figure}[t]     
\centerline{\resizebox{0.80\columnwidth}{!}{\includegraphics{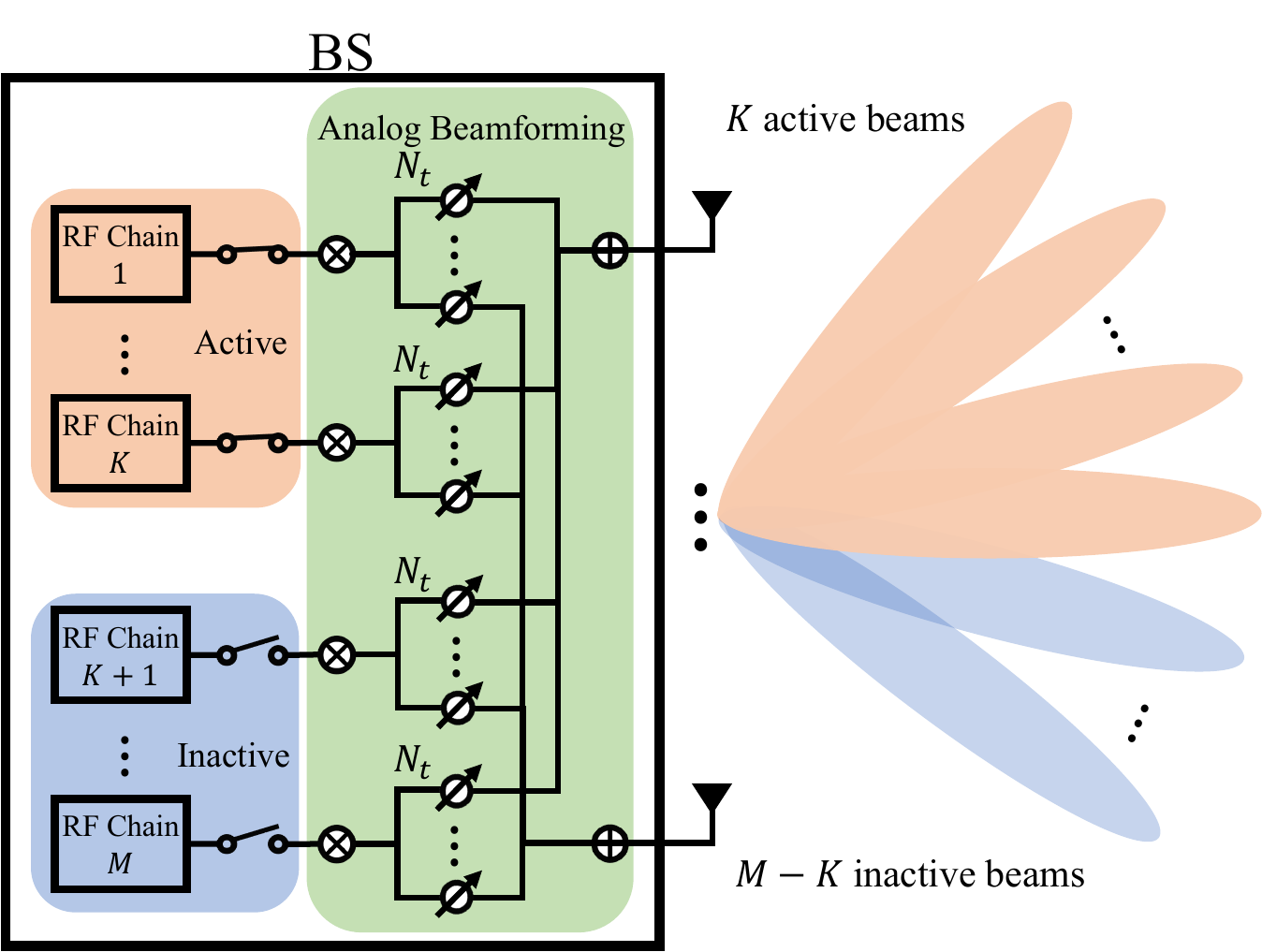}}}
     \caption{
     Beam selection with $K$ active RF chains.
     Orange beams represent the selected (active) beams, while blue beams represent unselected (inactive) beams.
     }\label{fig:beam}
\end{figure}

In the previous sections, we develop the GES and GCS algorithms under the assumption that each RF chain is connected to a single antenna.
In this section, we extend our proposed methods to the hybrid beamforming architecture \cite{heath:jstsp:16} with $M$ RF chains, each connected to all $N_t$ antennas via dedicated phase shifters as shown in Fig.~\ref{fig:beam}.
Since each RF chain then synthesizes a beamformed signal, i.e., a linear combination of the signals of all $N_t$ antennas, it corresponds to a distinct beam-pattern determined by its phase shifter configuration; accordingly, the original problem of RF chain selection naturally evolves into beam selection.
Assuming that analog beamforming is fixed and designed to divide the angular domain into $M$ evenly spaced beams \cite{amadori:tcom:15}, we represent the beamforming vector for the $m$-th RF chain as follows:
\begin{align}
    \mathbf{u}_{m} = \sqrt{\frac{1}{N_t}}\left[1, e^{-j2\pi \frac{m-1}{M}}, \ldots, e^{-j2\pi(N_t - 1) \frac{m-1}{M}} \right]^{\sf T} \in \mathbb{C}^{N_t}.
\end{align}
These beams are formed by uniformly selecting columns from the $N_t \times N_t$ unitary discrete Fourier transform (DFT) matrix, and we denote the analog beamforming matrix by $\mathbf{U} = [\mathbf{u}_1, \ldots, \mathbf{u}_M] \in \mathbb{C}^{N_t \times M}$.

With $\mathbf{U}$, the received signals \eqref{eq:comm_signal} and \eqref{eq:sensing_signal} are reformulated as
\begin{align}
    \mathbf{Y}_c &= \mathbf{\tilde H}_c \mathbf{\tilde X}+\mathbf{Z}_c,\\
    \mathbf{Y}_s^{\sf H} &= \mathbf{\tilde X}^{\sf H} \mathbf{\tilde H}_s^{\sf H} + \mathbf{Z}_s^{\sf H},
\end{align}
where
\begin{align}
    \mathbf{\tilde H}_c &= \mathbf{H}_c\mathbf{U} \in \mathbb{C}^{N_c \times M}, \\
    \mathbf{\tilde H}_s^{\sf H} &= \mathbf{U}^{\sf H} \mathbf{H}_s^{\sf H}  \in \mathbb{C}^{M \times N_s}.   
\end{align} 
Here, $\mathbf{\tilde X} \in \mathbb{C}^{M \times T}$ represents the signal from the RF chains, not the signals transmitted from the antennas, and $\mathbf{\tilde H}_c,\, \mathbf{\tilde H}_s^{\sf H}$ are the effective channel and TRM combined with the analog beamforming matrix, whose per-column covariance is $\mathbf{\tilde R}_{{\sf T},n} = \mathbf{U}^{\sf H}\mathbf{R}_{{\sf T}, n}\mathbf{U}$.

Accordingly, \eqref{eq:before_CMI} and \eqref{eq:SMI_each} are rewritten as
\begin{align}
    I_c(\CMcal{M}) &= T \log_2{\left| \mathbf{I}_{M} + \gamma \mathbf{\tilde H}_c^{\sf H}(\CMcal{M}) \mathbf{\tilde H}_c(\CMcal{M})\right|}, \label{eq:beamCMI}\\
    I_s(\CMcal{M}) &= \sum_{n=1}^{N_s} \log_2{\left| \mathbf{I}_{M} + \gamma T \mathbf{\tilde R}_{T,n}(\CMcal{M})  \right|} \label{eq:beamSMI}, 
\end{align} where $\CMcal{M}=\{1,\ldots, M\}$ is the entire RF chain set. 
Since the above equations have the same form as \eqref{eq:before_CMI} and \eqref{eq:SMI_each}, substituting $\mathbf{\tilde H}_c(\CMcal{M})$ and $\mathbf{\tilde R}_{{\sf T},n}(\CMcal{M})$ for $\mathbf{H}_c(\CMcal{N}_t)$ and $\mathbf{R}_{{\sf T}, n}(\CMcal{N}_t)$ therein and adjusting the size of the identity matrix keep the previously established theorems and corollaries applicable, so that the proposed GES and GCS methods can be used without loss of generality.



It is worth noting that under the following asymptotic conditions: i) $M \to N_t$, ii) $N_t \to \infty$ and $N_c \to \infty$, and iii) $\{\theta^c_\ell\}_{\ell=1}^L$ and $\{\phi^c_\ell\}_{\ell=1}^L$ are pairwise distinct,
the MI expressions \eqref{eq:beamCMI} and \eqref{eq:beamSMI} become significantly simplified.
Recalling \eqref{eq:full_commch} and \eqref{eq:sen_cov}, under these conditions $\{\mathbf{t}(\theta_{\ell}^c)\}_{\ell =1}^L$, $\{\mathbf{r}(\phi_{\ell}^c)\}_{\ell =1}^{L}$ and $\{\mathbf{t}(\theta_{n}^s)\}_{n =1}^{N_s}$ become orthogonal \cite{marzetta:twc:10} and $\{\mathbf{t}(\theta_{\ell}^c)\}_{\ell =1}^L$ and $\{\mathbf{t}(\theta_{n}^s)\}_{n =1}^{N_s}$ align closely with exactly one analog beamforming vector of $\mathbf{U}$ \cite{heath:jstsp:16}.
As a result, the columns of $\mathbf{\tilde H}_c(\CMcal{M})$ are nearly orthogonal, so that both $\mathbf{\tilde H}_c^{\sf H}(\CMcal{M})\mathbf{\tilde H}_c(\CMcal{M})$ and $\mathbf{\tilde R}_{T,n}(\CMcal{M})$ become approximately diagonal matrices.
Then, \eqref{eq:beamCMI} and \eqref{eq:beamSMI} are approximated as
\begin{align}
    I_c(\CMcal{M}) &\approx T \log_2{\left| \mathbf{\tilde D}(\CMcal{M})
    \right|} = \sum_{j=1}^M T \log_2{\tilde d_j}, \label{eq:dbs_CMI}\\
    I_s(\CMcal{M})
    &\approx \sum_{n=1}^{N_s} \log_2{\left|\mathbf{\tilde E}_n(\CMcal{M}) \right|} = \sum_{j=1}^{M} \sum_{n=1}^{N_s} \log_2{{\tilde e_{n,j}}}, \label{eq:dbs_SMI}
\end{align} where
\begin{align}
    \mathbf{\tilde D}(\CMcal{M}) &= \text{diag}\left(
    \mathbf{I}_{M} + \gamma \mathbf{\tilde H}_c^{\sf H}(\CMcal{M}) \mathbf{\tilde H}_c(\CMcal{M})
    \right) \nonumber \\
    &= \text{diag}([{\tilde d_1}, \ldots, {\tilde d_{M}}]^{\sf T}), \\
    \mathbf{\tilde E}_n(\CMcal{M}) &= \text{diag}\left(
    \mathbf{I}_{M} + \gamma T \mathbf{\tilde R}_{{\sf T},n}(\CMcal{M})
    \right) \nonumber \\ 
    &= \text{diag}([{\tilde e_{n,1}}, \ldots, {\tilde e_{n, M}}]^{\sf T}).
\end{align}

Note that \eqref{eq:dbs_CMI} and \eqref{eq:dbs_SMI} correspond to Theorem~\ref{eq_theo_gcs} under the asymptotic case, where $\delta_j$ and $\varepsilon_{n,j}$ reduce to
\begin{align}
\delta_j &= \left[ \mathbf{\tilde D}^{-1}(\CMcal{M}) \right]_{j,j} = \tilde d_j^{-1}, \\
\varepsilon_{n,j} &= \left[ \mathbf{\tilde E}^{-1} (\CMcal{M}) \right]_{j,j} = \tilde e_{n,j}^{-1}.
\end{align}
Hence, the contributions of the $j$-th RF chain to the communication and sensing MI are $T\log_2{\tilde d_j}$ and $\sum_{n=1}^{N_s} \log_2{\tilde e_{n,j}}$, respectively.
Unlike GES and GCS, where the contributions are computed via computationally expensive matrix inversions and thus require sequential updates (recall Corollaries~\ref{eq_cor_ges} and~\ref{eq_cor_gcs}), here they are obtained simply by extracting the diagonal entries of $\mathbf{\tilde D}(\CMcal{M})$ and $\mathbf{\tilde E}_n(\CMcal{M})$ and remain unchanged throughout the selection process.
As a result, it is possible to select the desired number of RF chains in a single step rather than sequentially, which further reduces the computational complexity.
We refer to this special case as diagonal beam selection (DBS), summarized in Algorithm~\ref{alg:DBS}; although developed under the aforementioned asymptotic conditions, DBS is shown in Section~\ref{sec_simul} to remain competitive even when these conditions are not satisfied.

\begin{algorithm}[t]
\caption{Diagonal Beam Selection} \label{alg:DBS} 
\textbf{input}: Desired number of RF chains $K$, weighting factors $\omega_c$, $\omega_s$ \\
Compute $\mathbf{\tilde D}$, $\mathbf{\tilde E}_n$  \\
Compute ${\tilde d_j}$, ${\tilde e_{n,j}}$ \\
Compute contributions
{\small
$
\left( \tilde d_j \right)^{\omega_c}  \left( \prod_{n=1}^{N_s} \tilde e_{n,j} \right)^{\frac{\omega_s}{N_s}}
$
} \\
Select $K$ RF chains with the highest contributions \\
\Return{$\CMcal{M} \leftarrow$}indices of selected RF chains
\end{algorithm}



\section{Numerical Results} \label{sec_simul}

We evaluate the proposed methods in the MIMO ISAC system with hybrid architecture, where a single RF chain is connected to $N_t$ antennas with particular phase shifters.
We first select $K_c$ and $K_s$ receive RF chains from $\CMcal{N}_c$ and $\CMcal{N}_s$, and then select $K$ transmit RF chains from $\CMcal{M}$.

As baselines for comparison, we consider the following methods.
{\textbf{Full-Selection}} uses all available RF chains at the BS, communication UE and sensing antennas without any selection process.
{\textbf{Exhaustive search (Exh)}} exhaustively searches all combinations of $K_c$ and $K_s$ receive RF chains from $\CMcal{N}_c$ and $\CMcal{N}_s$, respectively, and selects the one maximizing the weighted sum of normalized MI; then, given the selected receive RF chains, it selects $K$ transmit RF chains from $\CMcal{M}$ in the same manner.
{\textbf{Random selection (Random)}} randomly selects the transmit RF chains at the BS and the receive RF chains at both the communication UE and sensing antennas.
{\textbf{Fixed selection (Fixed)}} pre-determines the transmit and receive RF chains by uniformly sampling with fixed spacing and rounding; for example, when $K = 4$ and $M = 8$, the selected transmit RF chains are $\CMcal{M} = \{1, 3, 5, 7\}$.

The simulation follows the channel and TRM models of Section~\ref{sec_sys}, where the UE and the sensing targets are randomly distributed within the angular sector $[-\pi/3, \pi/3]$; that is, for the UE and each target located at angle $\theta$, the corresponding AoD is uniformly distributed over $[\theta-\pi/3, \theta+\pi/3]$.
Unless stated otherwise, the numbers of paths and targets are $L_p = L_t = 8$, the numbers of antennas and RF chains at the BS are $N_t = M = 16$, the numbers of RF chains at the UE and sensing side are $N_c = N_s = 8$, and the weights are $\omega_c = \omega_s = 0.5$; the remaining setups are given in the caption of each figure.


\subsection{Weighted Sum of Normalized MI}

\begin{figure}[t]     
\centerline{\resizebox{0.80\columnwidth}{!}
{\includegraphics{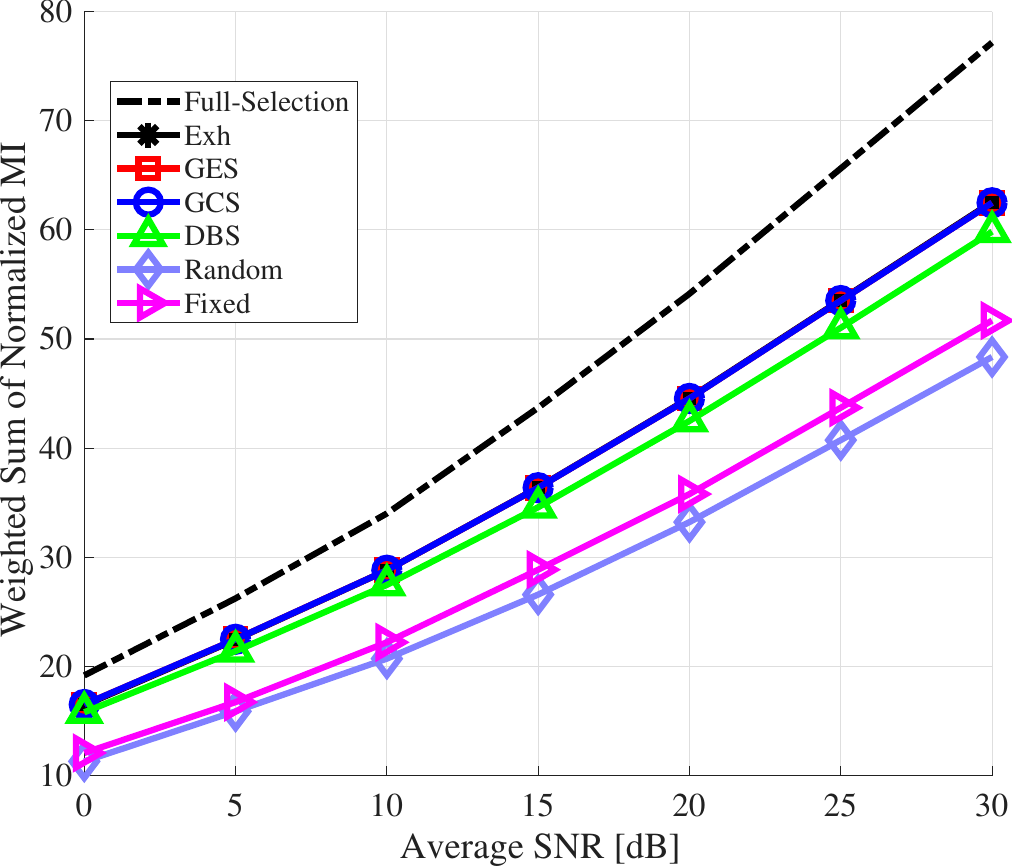}}}
     \caption{
The weighted sum of normalized MI over average SNR (dB) in beamspace, with $K = 8$ and $K_c = K_s = 6$.}
\label{sim_beamtxrx}
\end{figure}

\begin{figure}[t]     
\centerline{\resizebox{0.80\columnwidth}{!}{\includegraphics{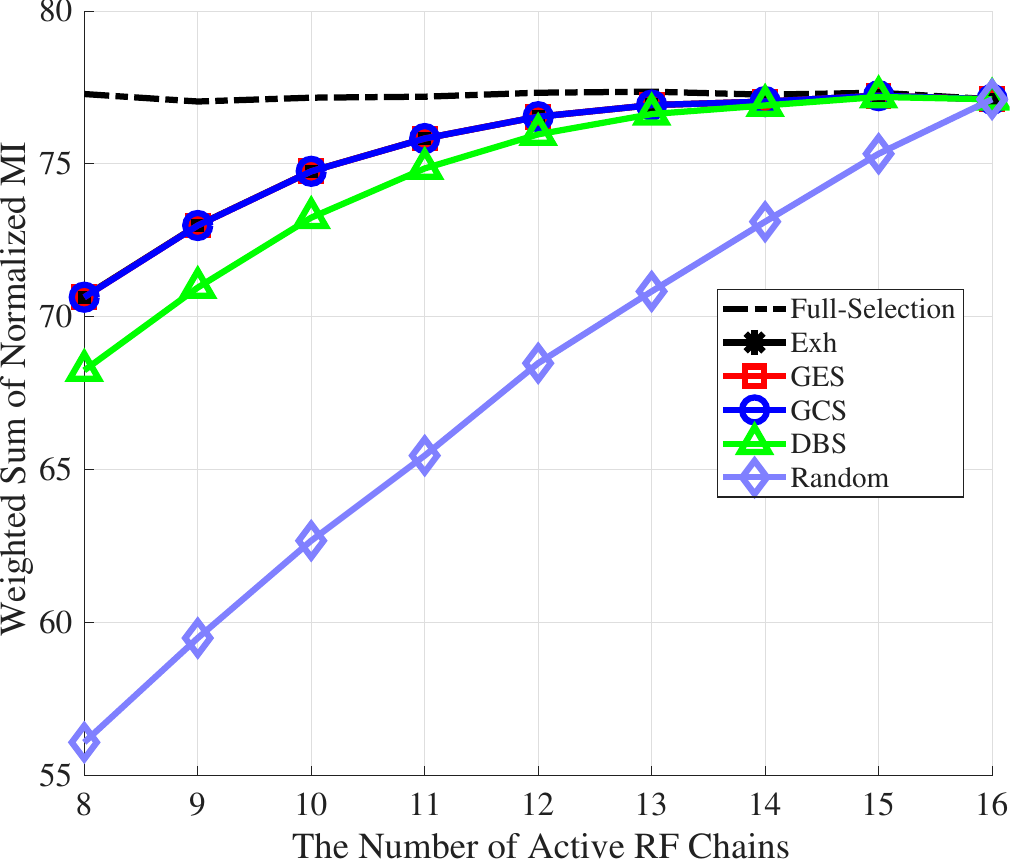}}}
     \caption{
The weighted sum of normalized MI over the number of active RF chains in beamspace, with $K_c = K_s = 8$ and average SNR of 30\,dB.}
     \label{sim_numRFmi}
\end{figure}


Fig.~\ref{sim_beamtxrx} illustrates the weighted sum of normalized MI as a function of the average SNR.
It is observed that GES and GCS exhibit performance nearly identical to Exh, while significantly outperforming Random and Fixed at an average SNR of 20\,dB by 32.15\% and 23.67\%, respectively.
To jointly enhance communication and sensing performance, RF chain selection must account for both the effective channel and the TRM; however, since each RF chain is intricately coupled with both factors, only a limited subset contributes significantly to both objectives, which in turn leads to highly imbalanced contributions to the total MI across the RF chains.
Accordingly, identifying RF chain subsets that account for the dominant contribution to the total MI is of critical importance, and the above observations demonstrate not only the compatibility of the proposed GES and GCS with the beamspace scenario, but also their effectiveness in capturing such imbalanced contributions.
DBS also performs very close to GES and GCS, with a gap of no more than 4.7\% even at 30\,dB, and thus remains competitive even when the asymptotic conditions are not satisfied.

Fig.~\ref{sim_numRFmi} illustrates the weighted sum of normalized MI as a function of the number of active RF chains. 
We focus on the regime where more than half of the RF chains are activated, so that the ISAC system sufficiently leverages the available spatial DoF.
Full-Selection remains unchanged with respect to the number of active RF chains, and Fixed, which uses a predetermined set regardless of this number, is excluded for clarity.
Again, GES and GCS achieve performance nearly identical to that of Exh, implying that the sequential reformulation of the RF chain selection problem incurs negligible performance loss.
We also note that DBS, despite its significantly reduced computational complexity, remains very close to GES and GCS, with the largest gap of only 3.57\% at 8 active RF chains, striking an excellent balance between performance and implementation efficiency.
In contrast, the Random exhibits poor performance except when activating nearly all RF chains, showing a 26.17\% performance loss compared to GES and GCS, which highlights the importance of an effective RF chain selection strategy across varying numbers of active RF chains.

Given the near-optimal performance of GES and GCS and the prohibitive computational burden associated with Exh, we exclude Exh from the subsequent simulations without loss of generality. 
\subsection{Energy Efficiency} \label{subsec_ee}
\begin{figure}[t]
\centerline{\resizebox{0.80\columnwidth}{!}
{\includegraphics{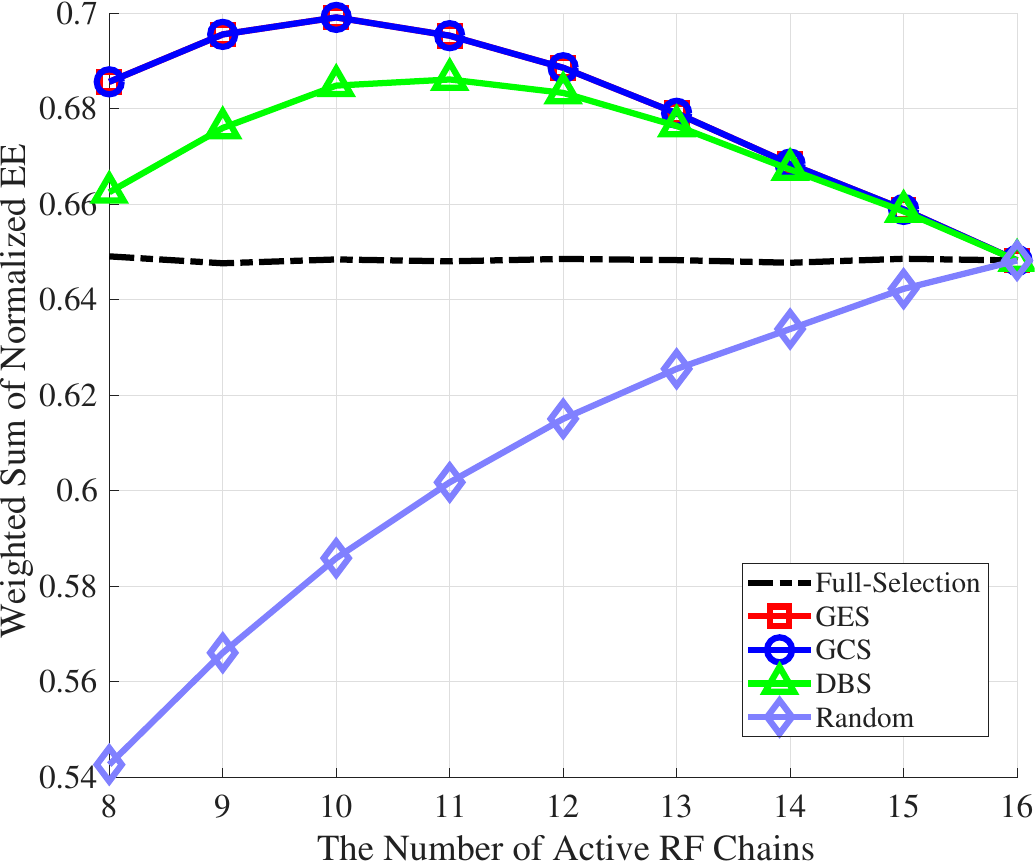}}}
\caption{
The weighted sum of normalized EE over the number of active RF chains in beamspace, with $K_c = K_s = 8$ and average SNR of 30\,dB.
All quantization resolutions are assumed to be 16 bits, i.e., $b_{{\sf DAC},n} = b_{{\sf ADC},n}^c = b_{{\sf ADC},n}^s = 16$.
}
\label{sim_numRFee}
\end{figure}

Now we examine energy efficiency (EE). To characterize normalized EE, we first consider total circuit power consumption as
\begin{align}
    P_{\sf cir}(\CMcal{N}_t, \CMcal{N}_c, \CMcal{N}_s) = P_{\sf BS}(\CMcal{N}_t) + P_{\sf UE}(\CMcal{N}_c) + P_{\sf s}(\CMcal{N}_s).
\end{align}
$P_{\sf BS}$, $P_{\sf UE}$, $P_{\sf s}$ are the circuit power consumption of the BS, communication UE, and sensing antennas, defined as
\begin{align}
    P_{\sf BS}(\CMcal{N}_t) &= 
    P_{\sf LO} + \sum_{n\in \CMcal{N}_t} \left( P_{\sf DAC}(b_{{\sf DAC},n}, f_s) + P_{\sf RF} \right), \\
    P_{\sf UE}(\CMcal{N}_c) &= P_{\sf LO} + \sum_{n\in \CMcal{N}_c} \left( P_{\sf ADC}(b_{{\sf ADC},n}^c, f_s) + P_{\sf RF} \right), \\
    P_{\sf s}(\CMcal{N}_s) &= P_{\sf LO} + \sum_{n\in \CMcal{N}_s} \left( P_{\sf ADC}(b_{{\sf ADC},n}^s, f_s) + P_{\sf RF} \right),
\end{align} where
$P_{\sf LO}$, $P_{\sf DAC}$, $P_{\sf ADC}$, $P_{\sf RF}$, $f_s$ denote the power consumption of local oscillator, digital-to-analog converter (DAC), analog-to-digital converter (ADC), RF chain, sampling rate, respectively.
$b_{{\sf DAC},n}$, $b_{{\sf ADC},n}^c$, $b_{{\sf ADC},n}^s$ denote quantization resolution (in bits) of $n$-th DAC at BS and $n$-th ADC at UE and sensing antennas, which are set to reflect a full-resolution hardware configuration.
All parameters and power consumption functions of the DAC and ADC follow \cite{choi:twc:22, shuguang:twc:05}.
Then, the normalized EE $\eta$ is defined as follows:
\begin{align}
    \eta(\CMcal{N}_t, \CMcal{N}_c, \CMcal{N}_s) = \frac{\frac{\omega_c}{T}I_c(\CMcal{N}_t) + \frac{\omega_s}{N_s}I_s(\CMcal{N}_t)}{P_{\sf cir}(\CMcal{N}_t, \CMcal{N}_c, \CMcal{N}_s)}.
\end{align} 

Fig.~\ref{sim_numRFee} illustrates the weighted sum of normalized EE $\eta$ of our proposed methods and baselines as a function of the number of active RF chains.
GES, GCS and DBS achieve better EE compared to Random, which is expected since they consume the same amount of power as Random while delivering significantly better performance, as shown in Fig.~\ref{sim_numRFmi}.
Notably, the proposed methods also outperform Full-Selection in terms of EE by up to 7.79\%, mainly because Full-Selection activates all RF chains and thus incurs significantly higher power consumption, which suggests that fully activating all RF chains does not necessarily lead to energy-efficient operation.

\subsection{Pareto Boundary of Normalized MI} \label{subsec_pareto}

\begin{figure}[t]     
\centerline{\resizebox{0.80\columnwidth}{!}{\includegraphics{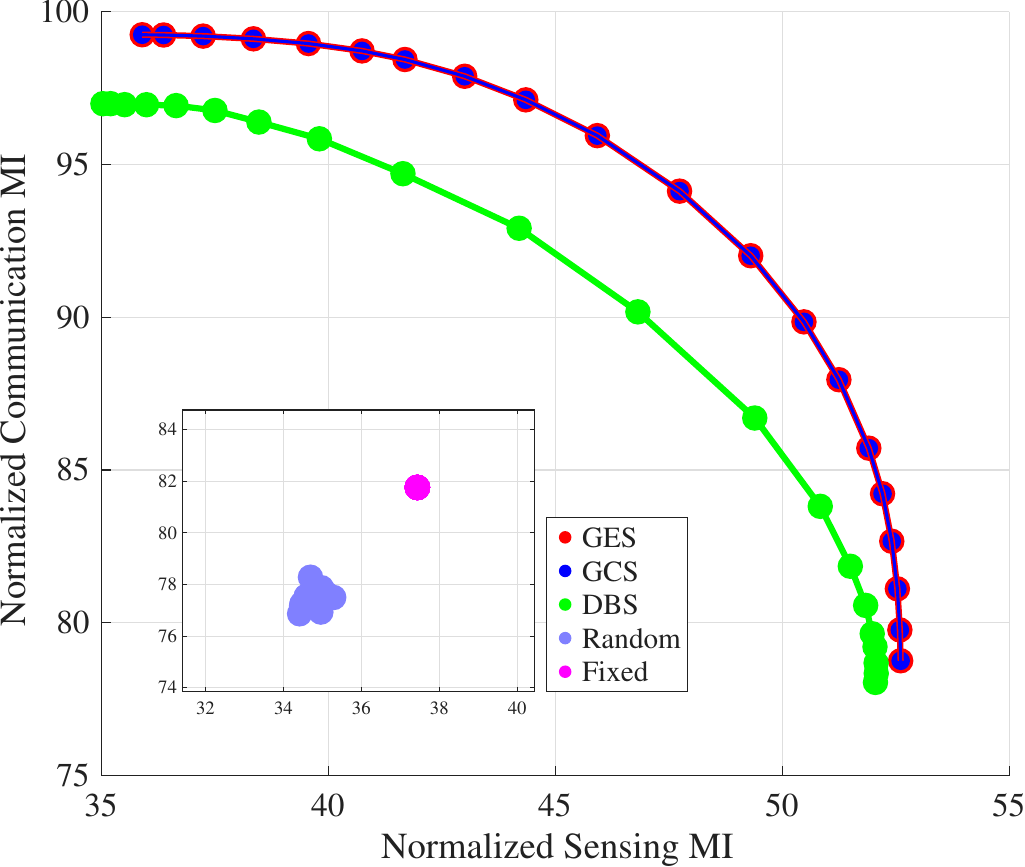}}}
     \caption{
The Pareto boundary of normalized communication MI and normalized sensing MI in beamspace, with $K = 8$ and average SNR of 30\,dB.   }
     \label{sim_beampareto}
\end{figure}
We plot the Pareto boundary of communication MI versus sensing MI as $\omega_c$ and $\omega_s$ vary in Fig.~\ref{sim_beampareto}, where receive RF chain selection is not considered for simplicity.
Each point on the curve represents a different weighting between communication and sensing objectives: the bottom right corresponds to $\omega_c=0$ (sensing-only), while the top left corresponds to $\omega_c=1$ (communication-only), with intermediate points reflecting gradual shifts in emphasis.

It is observed that the proposed GES and GCS methods achieve the best Pareto boundaries, whereas DBS, despite its performance comparable to GES and GCS in Fig.~\ref{sim_beamtxrx}, Fig.~\ref{sim_numRFmi} and Fig.~\ref{sim_numRFee}, yields a Pareto boundary relatively limited in extent.

This is explained as follows.
In GCS, the key factor that determines the contribution of each RF chain is the corresponding diagonal entry of the inverse matrix, which is influenced by the determinant of the matrix obtained by removing the corresponding row and column (see Appendix~\ref{appC}).
Since this determinant reflects the degree of orthogonality among the remaining channel and TRM vectors when the given RF chain is excluded, sequentially removing the RF chain associated with the largest such determinant is equivalent to eliminating the most redundant RF chain.
In contrast, DBS ignores the interdependence among RF chains captured by the off-diagonal entries and instead evaluates the contribution of each RF chain solely based on the power of its corresponding channel and TRM vectors.
As a result, DBS may favor high-power yet redundant RF chains that are already well represented in either the communication or sensing subspaces, which leads to a suboptimal selection that fundamentally limits the overall Pareto boundary performance of DBS.




These observations collectively suggest that, when a proper balance between communication and sensing is required, the Fixed and Random are unsuitable due to their inability to account for this trade-off, and that DBS, while better balanced than these two, still falls short of providing sufficiently diverse trade-off solutions.
In contrast, the proposed GES and GCS consistently achieve well-balanced and widely spread Pareto boundaries, demonstrating their effectiveness for joint communication and sensing optimization.

\subsection{Runtime Comparison of GES and GCS} \label{subsec_hmap}
\begin{figure}[t]
\centerline{\resizebox{0.80\columnwidth}{!}{\includegraphics{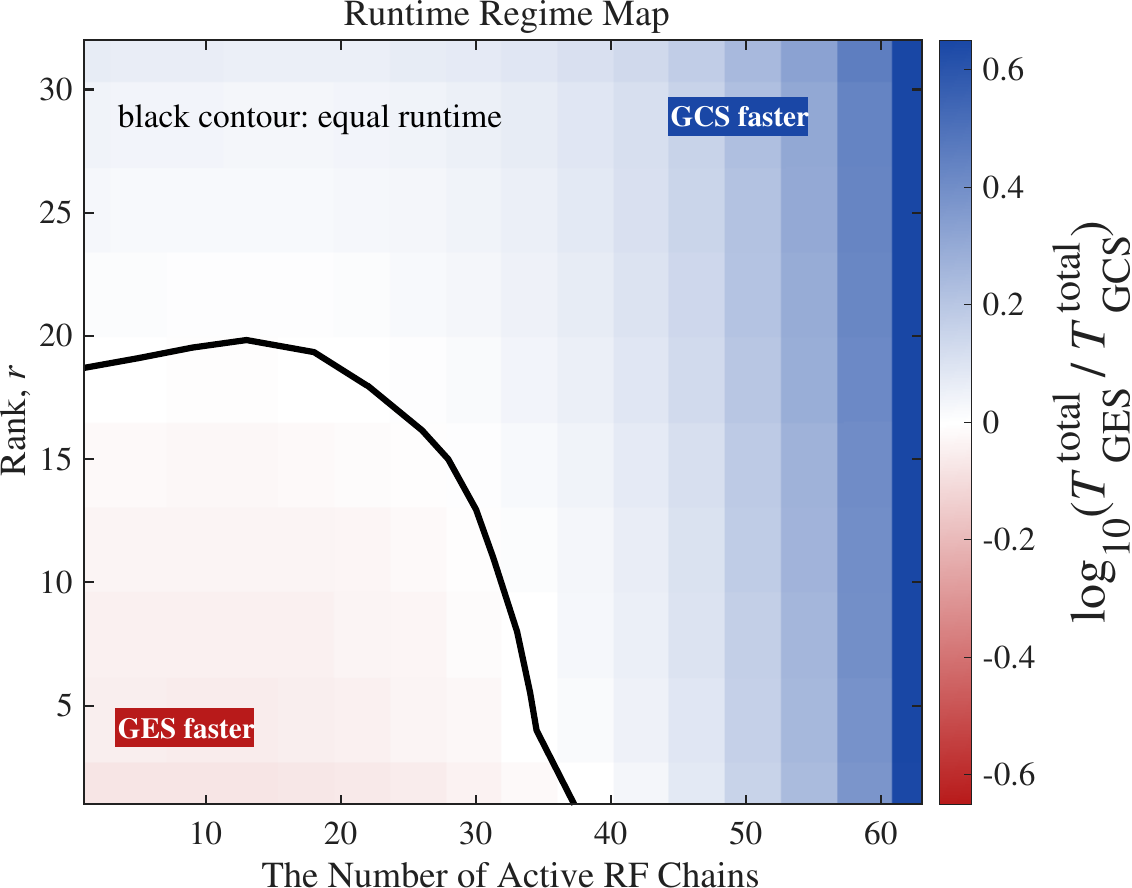}}}
\caption{Runtime regime map of the proposed GES and GCS over the
$(K,r)$ plane, showing $\log_{10}\!\big(T_{\rm GES}^{\rm total}/T_{\rm GCS}^{\rm total}\big)$, where $T_{(\cdot)}^{\rm total}$ is the total runtime averaged over $10^4$ realizations.
Red indicates that GES is faster, blue that GCS is faster, and the black contour marks equal runtime $T_{\rm GES}^{\rm total}=T_{\rm GCS}^{\rm total}$. Here, $N_t=64$ and $N_c=N_s=32$.\label{fig:hmap}}
\end{figure}

To validate the complementary computational behavior predicted in Section~\ref{subsec_GES_GCS}, we measure the runtime of GES and GCS over the sensing-rank/selection-depth plane, setting $r_n=r$ for all $n$ so that $r=\max_n r_n$ is the common sensing rank. The total runtime
$T_{\rm GES}^{\rm total}=T_{\rm GES}^{\rm init}+T_{\rm GES}^{\rm seq}$ and $T_{\rm GCS}^{\rm total}=T_{\rm GCS}^{\rm init}+T_{\rm GCS}^{\rm seq}$ is the sum of the initialization time and the sequential update time.
Each entry of Fig.~\ref{fig:hmap} reports $\log_{10}(T_{\rm GES}^{\rm total}/T_{\rm GCS}^{\rm total})$ averaged over $10^{4}$ realizations.

Fig.~\ref{fig:hmap} exhibits two clearly separated regimes divided by the equal-runtime contour.
In the low-rank and deep-selection region (small $r$ and small $K$, i.e., a large removal depth $N_t-K$), GES is faster, since its per-iteration saving in the rank domain, accumulated over the many removal iterations, outweighs its EVD-based initialization overhead.
As either $r$ increases or $K$ approaches $N_t$, the map turns blue and GCS becomes faster in the considered implementation because
its sequential cost is independent of $r$ and scales with the
remaining set size, while its initialization avoids EVD.
Hence, the runtime preference is jointly determined by the sensing rank $r$ and the selection depth $N_t-K$ rather than by either parameter alone.
We further observe that the selected sets of GES and GCS are identical in every tested realization, in agreement with Section~\ref{subsec_GES_GCS}; hence the runtime difference in Fig.~\ref{fig:hmap} arises purely from the computational realization rather than from different RF chain selections.
These results demonstrate that GES and GCS are two complementary realizations of the same MI-based greedy selection rule.

\section{Conclusion} \label{sec_concl}

In this paper, we proposed two low-complexity greedy RF chain selection methods for MIMO ISAC, by harnessing a unified MI performance metric for both communication and sensing, and revealed the following key findings.
First, by decomposing the MI into individual RF chain contributions, the proposed GES, GCS, and DBS algorithms enable efficient selection and achieve near-optimal performance, while significantly reducing computational complexity.
Second, despite Full-Selection achieving the highest MI, it suffers from poor EE due to its full hardware deployment, whereas the proposed methods achieve better EE by judiciously selecting a subset of RF chains.
Lastly, in the trade-off between communication and sensing, the proposed GES and GCS achieve the most extensive Pareto frontier, maintaining balanced performance across different weighting preferences.
These findings confirm that the proposed framework offers an effective and practical solution for RF chain selection in MIMO ISAC systems.

\appendices
\section{}\label{appA}
The communication MI with $\CMcal{\bar N}_t$ is represented as
\begin{align*}
    I_c(\CMcal{\bar N}_t) 
    &= T \log_2{\left| \mathbf{I}_{N_c} + \gamma \mathbf{H}_c(\CMcal{\bar N}_t) \mathbf{H}_c^{\sf H}(\CMcal{\bar N}_t)\right|}.
\end{align*}

When the $j$-th RF chain in $\CMcal{\bar N}_t = \{n_1,\ldots, n_{|\CMcal{\bar N}_t|}\}$ is removed, the corresponding column of $\mathbf{H}_c(\CMcal{\bar N}_t)$, denoted by $\mathbf{h}_j$, is also removed, so that the communication MI is represented as
\begin{align*}
    I_c(\CMcal{\bar N}_t \setminus \{j\})
    &= T \log_2{\left| \mathbf{I}_{N_c} + \gamma \left(\mathbf{H}_c(\CMcal{\bar N}_t) \mathbf{H}_c^{\sf H}(\CMcal{\bar N}_t) - \mathbf{h}_j \mathbf{h}_j^{\sf H}\right) \right|}, \\
    &= T \log_2{\left| \left(\mathbf{I}_{N_c} +\gamma \mathbf{H}_c(\CMcal{\bar N}_t) \mathbf{H}_c^{\sf H}(\CMcal{\bar N}_t) \right)
    - \gamma \mathbf{h}_j \mathbf{h}_j^{\sf H} \right|}.
\end{align*}
Applying the determinant lemma $|\mathbf{A}+\mathbf{u}\mathbf{v}^{\sf H}| = |\mathbf{A}|(1+\mathbf{v}^{\sf H} \mathbf{A}^{-1}\mathbf{u})$ decomposes $I_c(\CMcal{\bar N}_t \setminus \{j\})$ as
\begin{align*}
    &T \log_2{\left| \mathbf{I}_{N_c} +\gamma \mathbf{H}_c(\CMcal{\bar N}_t) \mathbf{H}_c^{\sf H}(\CMcal{\bar N}_t) \right|} \nonumber \\ 
    &- \underbrace{
    T \log_2{\left(\left(1 - \gamma \mathbf{h}_j^{\sf H}\left(\mathbf{I}_{N_c} +\gamma \mathbf{H}_c(\CMcal{\bar N}_t) \mathbf{H}_c^{\sf H}(\CMcal{\bar N}_t) \right)^{-1} \mathbf{h}_j \right)^{-1}\right) }
    }_{\text{The contribution of the $j$-th RF chain}} \\
    &= I_c(\CMcal{\bar N}_t) - T \log_2{\left( (1- \gamma \alpha_j)^{-1} \right)},
\end{align*} where
\begin{align*}
\alpha_j = \mathbf{h}_j^{\sf H} \mathbf{A} \mathbf{h}_j \text{ and } \mathbf{A} = \big( \mathbf{I}_{N_c} +\gamma \mathbf{H}_c(\CMcal{\bar N}_t) \mathbf{H}_c^{\sf H}(\CMcal{\bar N}_t) \big)^{-1}.    
\end{align*}

Applying EVD to $\mathbf{R}_{{\sf T}, n}$ yields
\begin{align*}
    \mathbf{R}_{{\sf T}, n} = \mathbf{\tilde P}_{n} \mathbf{\tilde \Lambda}_{n} \mathbf{\tilde P}_{n}^{\sf H}
    = \mathbf{P}_{n} \mathbf{\Lambda}_{n} \mathbf{P}_{n}^{\sf H}
    = \mathbf{G}_n \mathbf{G}_n^{\sf H},
\end{align*} where the columns of $\mathbf{\tilde P}_{n} \in \mathbb{C}^{N_t \times N_t}$ are eigenvectors of $\mathbf{R}_{{\sf T}, n}$ and the diagonal entries of $\mathbf{\tilde \Lambda}_n\in \mathbb{R}^{N_t \times N_t}$ are eigenvalues placed in descending order.
Supposing that the rank of $\mathbf{R}_{{\sf T}, n}$ is $r_n$, $\mathbf{R}_{{\sf T}, n}$ is decomposed into the submatrix $\mathbf{P}_n\in \mathbb{C}^{N_t \times r_n}$ of $\mathbf{\tilde P}_n$ and $\mathbf{\Lambda}_n\in \mathbb{R}^{r_n \times r_n}$ whose diagonal entries are the non-zero eigenvalues of $\mathbf{R}_{{\sf T}, n}$, and we define $\mathbf{G}_n = \mathbf{P}_{n} \mathbf{\Lambda}_{n}^{1/2} \in \mathbb{C}^{N_t \times r_n}$.

Then, the sensing MI $I_s(\CMcal{\bar N}_t)$ is expressed as
\begin{align*}
    I_s(\CMcal{\bar N}_t) &= \sum_{n=1}^{N_s} I_{s,n}(\CMcal{\bar N}_t)= \sum_{n=1}^{N_s} \log_2{\left| \mathbf{I}_{|\CMcal{\bar N}_t|} + \gamma T \mathbf{R}_{{\sf T}, n}(\CMcal{\bar N}_t) \right|}, \\
    &= \sum_{n=1}^{N_s} \log_2{\left| \mathbf{I}_{|\CMcal{\bar N}_t|} + \gamma T  \mathbf{G}_{n}(\CMcal{\bar N}_t) \mathbf{G}_{n}^{\sf H}(\CMcal{\bar N}_t)\right|}, \\ 
    &\mathop{=}^{(a)} \sum_{n=1}^{N_s} \log_2{\left| \mathbf{I}_{r_n} + \gamma T  \mathbf{G}_{n}^{\sf H}(\CMcal{\bar N}_t) \mathbf{G}_{n}(\CMcal{\bar N}_t)\right|},
\end{align*} where (a) follows from Weinstein-Aronszajn identity $|\mathbf{I}+\mathbf{A}\mathbf{B}| = |\mathbf{I}+\mathbf{B}\mathbf{A}|$, $\mathbf{G}_{n}(\CMcal{\bar N}_t)$ is $\mathbf{S}^{\sf T}(\CMcal{\bar N}_t)\mathbf{G}_{n}$, and $I_{s,n}(\CMcal{\bar N}_t)$ is the sensing MI between the received signal at the $n$-th sensing antenna and the $n$-th column of $\mathbf{H}_s^{\sf H}(\CMcal{\bar N}_t)$.

When the $j$-th RF chain of $\CMcal{\bar N}_t = \{n_1,\ldots,n_{|\CMcal{\bar N}_t|}\}$ is removed, the corresponding column of $\mathbf{G}_n^{\sf H}(\CMcal{\bar N}_t)$, denoted by $\mathbf{g}_{n,j}$, is also removed, so that the $n$-th sensing MI is formulated as
\begin{align*}
    I_{s, n}(\CMcal{\bar N}_t \setminus \{j\})= \log_2{\left| \mathbf{I}_{r_n} + \gamma T
    \left(
    \mathbf{G}_{n}^{\sf H}(\CMcal{\bar N}_t) \mathbf{G}_{n}(\CMcal{\bar N}_t)
    -\mathbf{g}_{n,j}\mathbf{g}_{n,j}^{\sf H}
    \right)
    \right|}.
\end{align*} 
Applying determinant lemma to the above equation yields
\begin{align*}
    &\log_2{\left| \mathbf{I}_{r_n} + \gamma T
    \mathbf{G}_{n}^{\sf H}(\CMcal{\bar N}_t) \mathbf{G}_{n}(\CMcal{\bar N}_t)
    \right|} \nonumber \\
    &\quad -
    \underbrace{
    \log_2{\left(  \left(1 - \gamma T\mathbf{g}_{n, j}^{\sf H}
    \left( \mathbf{I}_{r_n} + \gamma T
    \mathbf{G}_{n}^{\sf H}(\CMcal{\bar N}_t) \mathbf{G}_{n}(\CMcal{\bar N}_t) \right)^{-1}
    \mathbf{g}_{n,j}\right)^{-1} \right)}
    }_{\text{The $n$-th contribution of the $j$-th RF chain}} \\
    &= I_{s,n}(\CMcal{\bar N}_t)
    - \log_2{\left(  \left(1 - \gamma T \beta_{n, j}\right)^{-1} \right)},
\end{align*} where
\begin{align*}
    \beta_{n,j}= \mathbf{g}_{n, j}^{\sf H} \mathbf{B}_n \mathbf{g}_{n,j} \text{ and }
    \mathbf{B}_n = \left(  \mathbf{I}_{r_n} + \gamma T
    \mathbf{G}_{n}^{\sf H}(\CMcal{\bar N}_t) \mathbf{G}_{n}(\CMcal{\bar N}_t) \right)^{-1}.
\end{align*}
Therefore, total sensing MI $I_s(\CMcal{\bar N}_t \setminus \{j\})$ is decomposed as follows:
\begin{align*}
    I_s(\CMcal{\bar N}_t \setminus \{j\})
    &=
    I_{s}(\CMcal{\bar N}_t) 
    -
    \sum_{n =1}^{N_s} \log_2{\left(  \left(1 - \gamma T \beta_{n, j}\right)^{-1} \right)} .
\end{align*}

\section{} \label{appB}
For ease of exposition, we denote the $i$-th remaining RF chain set as $\CMcal{\bar N}_t^{(i)}$.
According to this notation, we additionally denote $\mathbf{H}_c^{(i)} = \mathbf{H}_c(\CMcal{\bar N}_t^{(i)}) = \mathbf{H}_c \mathbf{S}(\CMcal{\bar N}_t^{(i)})$ and $ \mathbf{G}_n^{(i)} = \mathbf{G}_n(\CMcal{\bar N}_t^{(i)}) = \mathbf{S}^{\sf T}(\CMcal{\bar N}_t^{(i)})\mathbf{G}_n $.

Assuming that the RF chain with the lowest contribution is removed, $\mathbf{A}^{(i+1)}$ is rewritten as 
\begin{align*}
    \mathbf{A}^{(i+1)} &= \left( \mathbf{I}_{N_c} +\gamma \mathbf{H}_c^{(i+1)} \mathbf{H}_c^{(i+1){\sf H}} \right)^{-1}\\
    &= \left( \underbrace{ \mathbf{I}_{N_c} +\gamma \mathbf{H}_c^{(i)} \mathbf{H}_c^{(i){\sf H}}}_{ (\mathbf{A}^{(i)})^{-1} } - 
    \gamma \mathbf{h}_{J^{(i)}} \mathbf{h}_{J^{(i)}}^{\sf H} \right)^{-1}.  
\end{align*}

Applying Woodbury matrix identity
$\left(\mathbf{A} + \mathbf{U}\mathbf{C}\mathbf{V} \right)^{-1}
= \mathbf{A}^{-1} - \mathbf{A}^{-1} \mathbf{U}
\left(\mathbf{C}^{-1} + \mathbf{V}\mathbf{A}^{-1}\mathbf{U} \right)^{-1}
\mathbf{V}\mathbf{A}^{-1}$ to the above equation
yields     
\begin{align*}
    &\mathbf{A}^{(i)} + \gamma \mathbf{A}^{(i)} \mathbf{h}_{J^{(i)}} 
    \left( 1- \gamma \mathbf{h}_{J^{(i)}}^{\sf H} \mathbf{A}^{(i)} \mathbf{h}_{J^{(i)}} \right)^{-1} \mathbf{h}_{J^{(i)}}^{\sf H} \mathbf{A}^{(i)}, \\
    &=
    \mathbf{A}^{(i)} + \gamma \left( 1- \gamma \alpha_{J^{(i)}}^{(i)} \right)^{-1} \mathbf{A}^{(i)} \mathbf{h}_{J^{(i)}} 
     \mathbf{h}_{J^{(i)}}^{\sf H} \mathbf{A}^{(i)}= 
    \mathbf{A}^{(i)} + \mathbf{a}^{(i)} \mathbf{a}^{(i) {\sf H}} ,
\end{align*} where $\mathbf{a}^{(i)} \triangleq \sqrt{\gamma\left( 1 - \gamma \alpha_{J^{(i)}}^{(i)} \right)^{-1}} \mathbf{A}^{(i)} \mathbf{h}_{J^{(i)}}$.

Note that, since $\mathbf{A}^{(i+1)}$ consists of $\mathbf{A}^{(i)}$ and $\mathbf{a}^{(i)}$ which are obtained at previous iteration, $\mathbf{A}^{(i+1)}$ can be sequentially updated easily.
For the same reason, $\alpha_j^{(i+1)}$ for all $j \in \CMcal{\bar N}_t^{(i+1)}$ are sequentially updated as 
\begin{align*}
    \alpha_j^{(i+1)} &= \mathbf{h}_j^{\sf H} \mathbf{A}^{(i+1)} \mathbf{h}_j = 
    \mathbf{h}_j^{\sf H} \left(
    \mathbf{A}^{(i)} + \mathbf{a}^{(i)}\mathbf{a}^{(i) {\sf H}}
    \right) \mathbf{h}_j, \\
    &= 
    \alpha_j^{(i)} + |\mathbf{h}_j^{\sf H}\mathbf{a}^{(i)}|^2.
\end{align*}

In the same way, $\mathbf{B}_n^{(i+1)}, \beta_{n,j}^{(i)}$ also can be sequentially updated as follows:
\begin{align*}
    &\mathbf{B}_n^{(i+1)} = \left( \mathbf{I}_{r_n} + \gamma T \mathbf{G}_{n}^{(i+1){\sf H}} \mathbf{G}_{n}^{(i+1)} \right)^{-1} \\ 
    &= \left( \underbrace{ \mathbf{I}_{r_n} + \gamma T \mathbf{G}_n^{(i) {\sf H}} \mathbf{G}_n^{(i)}}_{ (\mathbf{B}_n^{(i)})^{-1}} -\gamma T \mathbf{g}_{n,{J^{(i)}}} \mathbf{g}_{n,{J^{(i)}}}^{\sf H} \right)^{-1}, \\
    &=  
    \mathbf{B}_n^{(i)} +  \gamma T 
    \left( 1- \gamma T \beta_{n,{J^{(i)}}}^{(i)} \right)^{-1}
    \mathbf{B}_n^{(i)} \mathbf{g}_{n,{J^{(i)}}}\mathbf{g}_{n,{J^{(i)}}}^{\sf H} \mathbf{B}_n^{(i)},
    \\
    &= \mathbf{B}_n^{(i)} + \mathbf{b}_n^{(i)}\mathbf{b}_n^{(i) {\sf H}},
\end{align*} where $\mathbf{b}_n^{(i)} \triangleq \sqrt{ \gamma T \left( 1 - \gamma T\beta_{n, {J^{(i)}}}^{(i)} \right)^{-1}}\mathbf{B}_n^{(i)} \mathbf{g}_{n,{J^{(i)}}}$
and 
\begin{align*}
    \beta_{n,j}^{(i+1)} &= \mathbf{g}_{n,j}^{\sf H} \mathbf{B}_n^{(i+1)}\mathbf{g}_{n,j} = 
    \mathbf{g}_{n, j}^{\sf H} \left(
    \mathbf{B}_n^{(i)} + \mathbf{b}_n^{(i)}\mathbf{b}_n^{(i) {\sf H}}
    \right) \mathbf{g}_{n, j}, \\
    &= 
    \beta_{n,j}^{(i)} + |\mathbf{g}_{n,j}^{\sf H}\mathbf{b}_n^{(i)}|^2.
\end{align*}


\section{} \label{appC}
The communication MI with $\CMcal{\bar N}_t$ is represented as
\begin{align*}
    I_c(\CMcal{\bar N}_t) 
    = T \log_2{\left| \mathbf{I}_{|\CMcal{\bar N}_t|} + \gamma \mathbf{H}_c^{\sf H}(\CMcal{\bar N}_t) \mathbf{H}_c(\CMcal{\bar N}_t) \right|} = T \log_2{|\mathbf{D}(\CMcal{\bar N}_t)|},
\end{align*}
where we denote $\mathbf{I}_{|\CMcal{\bar N}_t|} + \gamma \mathbf{H}_c^{\sf H}(\CMcal{\bar N}_t) \mathbf{H}_c(\CMcal{\bar N}_t)$ by $\mathbf{D}(\CMcal{\bar N}_t)$.
When the $j$-th RF chain of $\CMcal{\bar N}_t = \{n_1,\ldots, n_{|\CMcal{\bar N}_t|}\}$ is removed, the corresponding row and column of $\mathbf{D}(\CMcal{\bar N}_t)$ are also removed, and this matrix is denoted by $\mathbf{D}(\CMcal{\bar N}_t \setminus \{j\})$.
By representing $\mathbf{D}^{-1}(\CMcal{\bar N}_t)$ in terms of the adjugate matrix of $\mathbf{D}(\CMcal{\bar N}_t)$ \cite{mtrxanalysis}, its $(j,j)$-th entry is expressed as
\begin{align*}
    [\mathbf{D}^{-1}(\CMcal{\bar N}_t)]_{j,j} = \frac{C_{jj}}{|\mathbf{D}(\CMcal{\bar N}_t)|} 
    = \frac{|\mathbf{D}(\CMcal{\bar N}_t \setminus \{j\})|}{|\mathbf{D}(\CMcal{\bar N}_t)|}, 
\end{align*} where $C_{jj} = (-1)^{(2j)}  |\mathbf{D}(\CMcal{\bar N}_t \setminus \{j\})|$ is cofactor of $[\mathbf{D}(\CMcal{\bar N}_t)]_{j,j}$. 
Therefore, we represent the communication MI with $\CMcal{\bar N}_t \setminus \{j\}$ as 
\begin{align*}
    &I_c(\CMcal{\bar N}_t \setminus \{j\}) = T \log_2{\left| \mathbf{D}(\CMcal{\bar N}_t \setminus \{j\})\right|}, \\
    &= T \log_2{\left| \mathbf{D}(\CMcal{\bar N}_t)\right|} 
    -\underbrace{T\log_2{ \left( [\mathbf{D}^{-1}(\CMcal{\bar N}_t)]_{j,j}^{-1}\right)} }_{\text{The contribution of the $j$-th RF chain}}, \\
    &= I_c(\CMcal{\bar N}_t) - T \log_2\left(\delta_j^{-1}\right),
\end{align*} where $\delta_j = [\mathbf{D}^{-1}(\CMcal{\bar N}_t)]_{j,j}$.

In the same way,
denoting $\left| \mathbf{I}_{|\CMcal{\bar N}_t|} + \gamma T \mathbf{R}_{{\sf T}, n}(\CMcal{\bar N}_t) \right|$ as $\mathbf{E}_n(\CMcal{\bar N}_t)$,
sensing MI with $\CMcal{\bar N}_t \setminus \{j\}$ is represented as 
\begin{align*}
    &I_s(\CMcal{\bar N}_t \setminus \{j\}) = \sum_{n=1}^{N_s} \log_2{\left| \mathbf{E}_n(\CMcal{\bar N}_t \setminus \{j\})\right|}, \\
    &= \sum_{n=1}^{N_s} \log_2{\left| \mathbf{E}_n(\CMcal{\bar N}_t)\right|} 
    -
    \underbrace{\sum_{n=1}^{N_s} \log_2{ \left( [\mathbf{E}_n^{-1}(\CMcal{\bar N}_t)]_{j,j}^{-1}\right)}}_{\text{The contribution of the $j$-th RF chain}}, \\
    &= I_s(\CMcal{\bar N}_t ) - \sum_{n=1}^{N_s} \log_2{ \left( \varepsilon_{n,j}^{-1} \right) }, 
\end{align*} where $\varepsilon_{n,j} = [\mathbf{E}_n^{-1}(\CMcal{\bar N}_t)]_{j,j}$.

\section{} \label{appD}
Denoting the $i$-th remaining RF chain set by $\CMcal{\bar N}_t^{(i)}$, we write $\mathbf{D}^{(i)} = \mathbf{D}(\CMcal{\bar N}_t^{(i)})$ and $ \mathbf{E}_n^{(i)} = \mathbf{E}_n(\CMcal{\bar N}_t^{(i)})$.
To sequentially update the inverses of $\mathbf{D}^{(i+1)}$ and $ \mathbf{E}_n^{(i+1)}$ via the Schur complement, we define the row permutation matrix
\begin{align*}
    [\mathbf{P}^{(i)}]_{k,:} =
\begin{cases}
\mathbf{e}_k^{\sf T}       & \text{for } 1 \leq k < J^{(i)} \\
\mathbf{e}_{k+1}^{\sf T}   & \text{for } J^{(i)} \leq k < |\CMcal{\bar N}_t^{(i)}| \\
\mathbf{e}_{J^{(i)}}^{\sf T}       & \text{for } k = |\CMcal{\bar N}_t^{(i)}|
\end{cases},
\end{align*}
where $ \mathbf{e}_k \in \mathbb{R}^{|\CMcal{\bar N}_t^{(i)}|} $ is the $k$-th standard basis vector in which the $k$-th entry is one and all other entries are zeros.
This matrix moves row $J^{(i)}$ to the last row and shifts rows $(J^{(i)}+1)$ through $|\CMcal{\bar N}_t^{(i)}|$ upward by one position; the corresponding column permutation is applied by multiplying $\mathbf{P}^{(i)\sf{T}}$ from the right.

Permuting both the rows and the columns of $\mathbf{D}^{(i)}$ by $\mathbf{P}^{(i)}$ and $\mathbf{P}^{(i) \sf{T}}$ places the RF chain to be removed in the last position, i.e.,
\begin{align*}
     \mathbf{\tilde D}^{(i)} \triangleq \mathbf{P}^{(i)}\mathbf{D}^{(i)}\mathbf{P}^{(i)\sf{T}} =
    \begin{bmatrix}
        \mathbf{D}^{(i+1)} & \mathbf{d}_{J^{(i)}} \\
        \mathbf{d}_{J^{(i)}}^{\sf H} & d_{J^{(i)}J^{(i)}}
    \end{bmatrix},
\end{align*}
so that the leading block of $\mathbf{\tilde D}^{(i)}$ is exactly $\mathbf{D}^{(i+1)}$ of the next iteration.
Since $\mathbf{P}^{(i)}$ is a permutation matrix, the inverse of $\mathbf{\tilde D}^{(i)}$ is obtained by applying the same permutation to $\mathbf{D}^{-(i)}$, i.e.,
\begin{align*}
\mathbf{\tilde D}^{-(i)} = \mathbf{P}^{(i)}\mathbf{D}^{-(i)}\mathbf{P}^{(i)\sf{T}} =
    \begin{bmatrix}
        \mathbf{\tilde D}_{11}^{-(i)} & \mathbf{\tilde D}_{12}^{-(i)} \\
        \mathbf{\tilde D}_{21}^{-(i)} & \mathbf{\tilde D}_{22}^{-(i)}
    \end{bmatrix},
\end{align*}
where the blocks are partitioned conformally with $\mathbf{\tilde D}^{(i)}$, i.e., $\mathbf{\tilde D}_{11}^{-(i)} \in \mathbb{C}^{|\CMcal{\bar N}_t^{(i+1)}| \times |\CMcal{\bar N}_t^{(i+1)}|}$ and $\mathbf{\tilde D}_{22}^{-(i)} \in \mathbb{C}$, and thus require no inversion operation.

By the block inversion formula, the leading block of the inverse of $\mathbf{\tilde D}^{-(i)}$, which is $\mathbf{\tilde D}^{(i)}$, equals the inverse of the Schur complement of $\mathbf{\tilde D}_{22}^{-(i)}$. Since this leading block is $\mathbf{D}^{(i+1)}$, we obtain
\begin{align*}
\mathbf{D}^{(i+1)} = \left( \mathbf{\tilde D}^{-(i)}/\mathbf{\tilde D}_{22}^{-(i)} \right)^{-1}
    = \left( \mathbf{\tilde D}_{11}^{-(i)} - \frac{\mathbf{\tilde D}_{12}^{-(i)}\mathbf{\tilde D}_{21}^{-(i)} }{ \mathbf{\tilde D}_{22}^{-(i)} } \right)^{-1},
\end{align*}
and taking the inverse of both sides yields
\begin{align*}
\mathbf{D}^{-(i+1)}
    = \mathbf{\tilde D}_{11}^{-(i)} - \frac{\mathbf{\tilde D}_{12}^{-(i)}\mathbf{\tilde D}_{21}^{-(i)} }{ \mathbf{\tilde D}_{22}^{-(i)} }.
\end{align*}
Since $\mathbf{\tilde D}_{11}^{-(i)}$, $\mathbf{\tilde D}_{12}^{-(i)}$, $\mathbf{\tilde D}_{21}^{-(i)}$, and $\mathbf{\tilde D}_{22}^{-(i)}$ are already obtained by permuting on both rows and columns of $\mathbf{D}^{-(i)}$,
we sequentially update $\mathbf{D}^{-(i+1)}$ without the inversion operation.

In the same way, $\mathbf{E}_n^{(i+1)}$, $\mathbf{E}_n^{-(i+1)}$ are obtained as 
\begin{align*}
    &\mathbf{\tilde E}_n^{(i)} \triangleq \mathbf{P}^{(i)}\mathbf{E}_n^{(i)}\mathbf{P}^{(i){\sf T}} = 
    \begin{bmatrix}
        \mathbf{E}_n^{(i+1)} & \mathbf{f}_{n, {J^{(i)}}} \\
        \mathbf{f}_{n, {J^{(i)}}}^{\sf H} & e_{n, J^{(i)}J^{(i)}}
    \end{bmatrix}, \\
    &\mathbf{\tilde E}_n^{-(i)} = \mathbf{P}^{(i)}\mathbf{E}_n^{-(i)}\mathbf{P}^{(i){\sf T}} = 
    \begin{bmatrix}
        \mathbf{\tilde E}_{n 11}^{-(i)} & \mathbf{\tilde E}_{n12}^{-(i)} \\
        \mathbf{\tilde E}_{n21}^{-(i)} & \mathbf{\tilde E}_{n22}^{-(i)}
    \end{bmatrix}, \\
    &\mathbf{E}_n^{-(i+1)} = \mathbf{\tilde E}_{n11}^{-(i)} - \frac{\mathbf{\tilde E}_{n12}^{-(i)}\mathbf{\tilde E}_{n21}^{-(i)} }{ \mathbf{\tilde E}_{n22}^{-(i)} }.
\end{align*}

\bibliographystyle{IEEEtran}
\bibliography{ref_isac_antsel}

\end{document}